\documentclass[reqno]{amsart}

\usepackage[letterpaper,margin=1.0in]{geometry}
\usepackage[english]{babel}
\usepackage{amsmath}
\usepackage{amssymb}
\usepackage{eucal}
\usepackage{amsthm}
\usepackage{graphicx,color}
\usepackage{setspace}
\usepackage{mathrsfs}
\usepackage{makecell}
\usepackage[titletoc,title]{appendix}
\usepackage{mathtools}
\usepackage{cite}
\usepackage{multicol}

\usepackage{color}
\usepackage{amsaddr}

\newcommand{\wa}{{{\color{white}a}}}
\newcommand{\wg}{{{\color{white}g}}}

\numberwithin{equation}{section}

\newcommand{\ed}[2]{\frac{\partial}{\partial #1}\text{#2}}
\newcommand{\edt}[2]{\frac{\partial^2}{\partial #1^2}\text{#2}}
\newcommand{\edm}[3]{\frac{\partial^2}{\partial #1\partial #2}\text{#3}}

\newcommand{\tder}{\frac{\partial}{\partial t}}
\newcommand{\tdd}{\frac{\partial^2}{\partial t^2}}
\newcommand{\rd}{\frac{\partial}{\partial r}}
\newcommand{\rdd}{\frac{\partial^2}{\partial r^2}}
\newcommand{\dtdr}{\frac{\partial^2}{\partial t\partial r}}

\newcommand{\tA}{\text{A}}
\newcommand{\tB}{\text{B}}
\newcommand{\tC}{\text{C}}
\newcommand{\tD}{\text{D}}
\newcommand{\tE}{\text{E}}
\newcommand{\tF}{\text{F}}
\newcommand{\tG}{\text{G}}
\newcommand{\tH}{\text{H}}
\newcommand{\tJ}{\text{J}}
\newcommand{\tK}{\text{K}}

\newcommand{\tP}{\text{P}}
\newcommand{\tQ}{\text{Q}}
\newcommand{\tR}{\text{R}}
\newcommand{\tS}{\text{S}}

\title{Gauge invariant perturbations of the Schwarzschild spacetime}
\author[Jonathan E. Thompson, Hector Chen and Bernard F. Whiting]
{Jonathan E. Thompson$^\dagger$, Hector Chen$^\dagger$ and  Bernard F. Whiting$^\dagger$} \email{thompson.jon@ufl.edu}
\email{hectorchen@ufl.edu}
\email{bernard@phys.ufl.edu}
\address{$^\dagger$Department of Physics, University of Florida, 2001 Museum Road, Gainesville, FL 32611-8440, USA}

\begin{document}


\begin{abstract}
Beginning with the pioneering work of Regge and Wheeler \cite{RW}, there have been many studies of perturbations away from the Schwarzschild spacetime background. In particular several authors \cite{Moncrief,Sachs,brizuela} have investigated gauge invariant quantities of the Regge-Wheeler (RW) formalism. Steven Detweiler also investigated perturbations of Schwarzschild in his own formalism, introducing his own gauge choice which he denoted the ``easy (EZ) gauge", and which he was in the process of adapting for use in the second-order self-force problem. We present here a compilation of some of his working results, arising from notes for which there seems to have been no manuscript in preparation.  In particular, we outline Detweiler's formalism, list the gauge invariant quantities he used, and explain the process by which he found them.
\end{abstract}

\maketitle

\section{Introduction}
\label{sec:intro}

The gravitational self-force problem has probably lead to one of the most intensive, long term, programs making use of black hole perturbation theory, possibly even surpassing in scale the application of quantum fields in curved (non-cosmological) spacetimes.  As such, it has lead to a number of computational (both numerical and analytical) refinements derived for efficiency and effectiveness.  In this paper we identify and catalogue an extended set of gauge invariant perturbations of the Schwarzschild geometry suitable for use in any perturbative analysis.  This list is based on notes originally developed by Steven L Detweiler (September 4th, 1947 to February 8th, 2016), and represents a core for the inner working of papers he has published on the self-force problem over the last fifteen years. Those rudimentary notes have been subsequently developed and extended here for publication, and we hope they will be of particular benefit to the wider self-force community.

The outline of the paper is as follows.  In section \ref{sec:pertintro} we give a very brief introduction to the foundations of perturbation theory and its relationship to gauge freedom.  In section \ref{sec:gaugeinvintro} we indicate how the notion of gauge invariance has been thought about and explored, and how the work we describe below can be viewed as fitting into that framework.  We are mindful of the fact that this may not represent the way Steven Detweiler might have talked or written about the topic, but we hope it provides a bridge between the formal and the very practical.  The beginning of Detweiler's personal approach is laid out in section \ref{sec:harmonicbasis}, starting with his tensor basis.  In section \ref{sec:a-knotation} we introduce the specific internal, A--K notation that Detweiler used and, for ease of translation into something more familiar, we relate his notation, in both the even- and odd-parity sectors, to that of Regge and Wheeler\cite{RW}.  Following that, in section \ref{sec:gaugeina-k}, we introduce Detweiler's characterization of a gauge vector, and explicitly give the vector required to go from any arbitrary gauge to either the Regge-Wheeler (RW) or Detweiler's own easy (EZ) gauge.  In section \ref{sec:gaugeinva-k} we briefly outline Detweiler's method.  We then derive his six fundamental gauge invariants (for $\ell\ge2$), four of even-parity and two of odd-parity, and list an additional one in each sector that can be related to the others. In section \ref{sec:metricrecon} we show how the full metric can be reconstructed from components of the Einstein tensor (effectively, source components of the Stress-Energy tensor) for the case $\ell\ge2$, using the EZ gauge as an example. 

In section \ref{GI:l=0,1}, we apply Detweiler's method to investigate gauge invariants for $\ell=0$ and $1$.  We pay particular attention to the special cases of $\ell=0$ and $\ell=1$, since these play an important role, especially for self-force regularization to be carried out correctly.
Using the example of a point particle in circular orbit, in section \ref{sec:simpex} we also show how to completely solve for the $\ell=0$ and $1$ metric components and determine the global changes in mass and angular momentum of the system, and sketch an outline of the solution method for \(\l\ge2\).  We conclude with a brief discussion (section \ref{sec:disc}), followed by three appendices, which respectively i) fully relate A--K to the Regge-Wheeler variables by integration over the two-sphere (appendix \ref{sec:a-kdecomp}), ii) express the Einstein tensor components in terms of A--K (appendix \ref{sec:einsteina-k}), and iii) express the Einstein tensor components solely in terms of Detweiler's gauge invariants.

Throughout this manuscript we choose to use Schwarzschild coordinates \((t,r,\theta,\phi)\) on the background spacetime and work in geometrized units \(c=G=1\). Lower case latin indices are used as spacetime indices and run from 0 to 3. The background metric is labeled by \(g^0_{ab}\) with signature \((-,+,+,+)\) and is used to raise and lower spacetime indices.

\section{Perturbation Theory and Gauge Freedom}
\label{sec:pertintro}

Exact solutions to the Einstein equations of General Relativity are few in number.  Thus, the direction of much progress has been in considering small perturbations of known solutions. In perturbation theory we start with an exact solution \((M_0,g_0)\) and search for a new solution \((M, g)\). The characterizing quality that makes \((M_0,g_0)\) an `exact solution' is the fact that we have in hand an atlas of coordinate charts (e.g. Schwarzschild coordinates for the Schwarzschild solution) which give us some idea of how various quantities behave on the manifold \(M_0\). What we would like to do is use these same coordinates on the physical manifold \(M\). These coordinate charts currently only map from (subsets of) \(\mathbb{R}^4\) to \(M_0\) and we would like to map them to \(M\) as well. We can accomplish this with a map \(\phi\): \(M_0\) \(\rightarrow\) \(M\). Since we may wish to extend a number of smoothly related coordinates, it is natural to demand that \(\phi\) also be a smooth map. Finally, we wish every point in \(M\) to have a coordinate labeling, but that no two points have the same coordinates. Thus we wish for \(\phi\) to be a bijection. A smooth bijection between manifolds is called a {\it diffeomorphism}. Given a coordinate system on the background manifold \(M_0\), a diffeomorphism assigns the same coordinate values smoothly among the points of the physical manifold \(M\). In situations where we expect the physical spacetime \((M,g)\) to differ only slightly from the background spacetime \((M_0,g_0)\), \(\phi\) simply tells us which points are to be considered `the same point'.  A diffeomorphism allows us to compare tensors at different points with the same coordinate values. For example.
\begin{align}
h_\phi \coloneqq \phi^*g -g_0,
\label{hdef}
\end{align}
is a tensor field on \(M_0\) colloquially known as the perturbation of the metric. The subscript \(\phi\) is used to emphasize that the value of \(h\) depends on the particular map used. The symbol $\phi^*$ is the pullback and denotes that the components of $g$ are transformed to the coordinates induced by $\phi$ in the usual way. Note that the pullback \(\phi^*g\) is also dependent on \(\phi\).

There is no universally preferred mapping \(\phi\) between any two manifolds and the value of any tensor or tensor perturbation depends in general on the particular correspondence between \(M_0\) and \(M\). Since we have considerable freedom to choose the mapping without changing the physical situation, it is appropriate to call a choice of \(\phi\) a {\it gauge} choice. Since a choice of $\phi$ causes a corresponding change in the perturbation in Eq.~\eqref{hdef} it is acceptable, and often more practical, to associate a given perturbation \(h_\phi\) with a gauge choice. We have the ability to choose any gauge so long as all equations are written in terms of physical quantities that do not depend on the choice of gauge. This constitutes what is known as the \textit{gauge freedom of perturbation theory}. The gauge independent physical quantities are known as \textit{gauge invariants}. 

\section{Gauge Invariance}
\label{sec:gaugeinvintro}

Let us now work in the perturbative sense in which \((M_0,g_0)\) and \((M,g)\) differ only slightly so that  Eq.~\eqref{hdef} is small everywhere for at least one gauge \(\phi\). Of course there is no guarantee that Eq.~\eqref{hdef} will remain small in a different gauge \(\psi\), for \(\psi^{-1}\) could map to a point where \(g_0\) is very different. The fact that different gauges reference different points of the background manifold is formally expressed by
\begin{align} \label{gaugetrans2}
\nonumber \chi: M_0 \rightarrow M_0,  \\
 \chi=\psi^{-1} \circ \phi. 
 \end{align}
Note that $\chi$ is a diffeomorphism from the background manifold to itself, and that $\phi \circ \chi^{-1}  = \psi$. The diffeomorphism $\chi$ can then be used to change from one gauge to another. From Eq.~\eqref{hdef} there will be a corresponding change in the perturbation:
\begin{align}
h_\psi = \psi^*g - g_0 = \chi_*\phi^*g-g_0.
\label{gaugetrans} \end{align}

Eqs.~\eqref{hdef} and \eqref{gaugetrans} must be compared at the same point. This is achieved by applying the pushforward $\chi_*$ to $h_\phi$, which results in
\begin{align}
\Delta h_\chi \equiv h_\psi - \chi_*h_\phi = \chi_*g_0 - g_0,
\label{gaugetrans3}\end{align}
an expression that depends only on $\chi$ and the background metric. Eq.~\eqref{gaugetrans3} formally expresses how the perturbation on the background spacetime changes due to a gauge transformation. We would like to examine when this change is small, so that the perturbation remains small. Since $g_0$ is a smooth tensor field, the right side of Eq.~\eqref{gaugetrans3} will be small when $\chi$ maps to an infinitesimally close point. Infinitesimal diffeomorphisms are generated by vector fields. Suppose $\chi$ is an infinitesimal diffeomorphism generated by a vector field $\xi^a$. The change of any tensor field under an infinitesimal pushforward $\chi_*$ is the Lie derivative of that tensor field with respect to $\xi^a$. In particular the change of the perturbation in Eq.~\eqref{gaugetrans3} will be
\begin{align}
\Delta h_\chi =  \pounds_\xi g_0 = \nabla_a\xi_b + \nabla_b\xi_a,
\label{gaugetrans4} \end{align}
where \(\xi\) is called a \textit{gauge vector}. Since this expression is not zero for an arbitrary vector field, the components of $h_{ab}$ vary under a choice of gauge. Since we expect that any interesting physical quantity is not dependent on a choice of gauge, our goal is to find quantities which remain the same under such transformations. This is the notion of \textit{gauge invariance} which we now explore.
 
At first it may be tempting to think of gauge invariant quantities as generic tensor fields \(T\) on \(M\) such that \(\phi^*T\) is the same for all gauges \(\phi\). While this seems to be a very natural definition of a gauge invariant quantity, careful analysis reveals some issues with it \cite{Sachs}\cite{Bruni}. Firstly, it is far too restrictive on the quantities T, which must be either vanishing tensors or constant scalar fields \cite{Sachs}. Secondly, the quantity \(\phi^*T\) is a tensor field \textit{on the background \(M_0\)}. If we want gauge invariants to be the interesting, measurable physical quantities in our theory, then we care about their values \textit{in the physical spacetime \(M\)}. Whether or not their value in the abstract background is independent of gauge is of no concern. Furthermore, in perturbation theory, the quantities of interest are formed out of perturbations of tensor fields, rather than the tensor fields themselves. The gauge invariance of perturbations was first examined by Stewart and Walker \cite{Stewart} who found that the perturbation \(\Delta T\) of a quantity \(T\) (both on the physical manifold) is gauge invariant if the value of the quantity itself is zero on the background, \(T_0=0\).
 
Let us now restrict our discussion of gauge invariance to perturbations of the metric tensor. From the definition of a gauge and Eq.~\eqref{hdef}, it is evident that we must always be working in some gauge when talking about \(h_{ab}\). Given an arbitrary choice of gauge, the metric perturbation \(h_{ab}\) is not likely to have any desirable properties. There exist a handful gauge choices in which \(h_{ab}\) does exhibit simple properties. Examples include the RW gauge which sets four components of \(h_{ab}\) to zero and Detweiler's EZ gauge which puts a slightly different set of components to zero (see Sec. \ref{subsec:EZgauge}). To find gauge invariants the idea is to start in an arbitrary gauge \(\psi\) and show the existence of the corresponding gauge vector \(\xi\) which transforms the tensor perturbation to one of these useful gauges according to Eq.~\eqref{gaugetrans4}. In particular the freedom in the arbitrary gauge vector must be used entirely to set the desired components of the metric (and only them) to their restricted values.

Historically, this method of computing gauge invariants was employed by Bardeen~\cite{Bardeen} and others. The examples in the body of this paper proceed slightly more generally. The left hand side of Eq.~\eqref{gaugetrans4} is evaluated under the transformation from an arbitrary gauge to another arbitrary gauge. The equation will then express how the components of the metric perturbation change in terms of an arbitrary gauge vector. In general none of the components of \(h_{ab}\) will be found to remain constant under a gauge transformation (see Eq.~\eqref{gaugetrans4}), but some combination of them might. 

Several authors have previously analyzed the issue of gauge invariant quantities. Moncrief \cite{Moncrief} analyzes perturbations variables from a variational principle that is independent of gauge. Gerlach and Sengupta \cite{Gerlach} use a $2+2$ split of spacetime to find gauge invariant quantities closely related to those in the RW gauge. Anderson \textit{et al.} \cite{Anderson} find gauge invariant quantities related to time derivatives of the extrinsic curvature of spacelike hypersurfaces in the Schwarzschild geometry. Jezierski \cite{Jezierski} introduces gauge invariant metric combinations similar to Gerlach and Sengupta, however without the use of a multipole decomposition, and associates these gauge invariants with the energy and angular momentum of gravitational waves. Gleiser \textit{et al.} \cite{Gleiser} give a systematic review of gauge transformations mainly focused on the RW and Zerilli results. They show that the RW variables (the remaining nonzero metric components in the RW gauge) are indeed gauge invariant. Finally, Martel and Poisson \cite{MartelPoisson} present a covariant formulation of the RW gauge invariants. The analysis in this paper follows in a similar fashion to that of Regge and Wheeler, however we will also focus specifically on the low-multipole sectors, and discuss the existence of gauge invariant quantities therein. In the rest of the manuscript we present an  application of the ideas discussed above to find gauge invariant quantities of the perturbed Schwarzschild spacetime. 

\section{Orthogonal Tensor Basis}
\label{sec:harmonicbasis}

The practical application of the approach we outline begins by establishing a basis, which is used to decompose tensor fields on the background metric. We choose our basis to be orthogonal, and utilize the spherical symmetry of the Schwarzschild spacetime to construct our basis out of scalar spherical harmonics and pure-spin vector and tensor harmonics.

To begin, two unnormalized, constant and orthogonal co-vector fields \(v\) and \(n\) are defined, with components in Schwarzschild coordinates of,
\begin{align*}
v_a&=(-1,0,0,0),&n_a&=(0,1,0,0),
\end{align*}
along with the projection operator onto the 2-sphere,
\begin{align*}
\sigma_{ab}&\equiv g^0_{ab}- \left(1-\frac{2M}r\right)^{-1} n_an_b+\left(1-\frac{2M}{r}\right) v_av_b\\
&=r^2\;\mathrm{diag}(0,0,1,\sin^2\theta),
\end{align*}
and the spatial Levi-Civita tensor, \(\epsilon_{abc}\equiv v^{d}\epsilon_{dabc}\), with \(\epsilon_{tr\theta\phi}=r^2\sin\theta\).

The pure-spin vector harmonics used by Detweiler are adapted from Thorne~\cite{Thorne} but with a different normalization:
\begin{align*}
Y^{E,\ell m}_{a}&=r\nabla_aY^{\ell m},&Y^{B,\ell m}_{a}&=r\epsilon_{ab}^{\;\;\;c}n^b\nabla_cY^{\ell m},&Y^{R,\ell m}_{a}&=n_aY^{\ell m},
\end{align*}
with \(Y^{\ell m}=Y^{\ell m}(\theta,\phi)\) being the scalar spherical harmonic,
\[
\oint Y^{\ell m}\left(Y^{\ell' m'}\right)^*\mathrm{d}\Omega=\delta_{\ell \ell'}\delta_{m m'}.
\] 
Here, the integration is performed over the 2-sphere, \(\mathrm{d}\Omega=\sin\theta\mathrm{d}\theta\mathrm\phi\), and an asterisk denotes complex conjugation. The vector harmonics above are mutually orthogonal and the integration will appear as
\[
\oint Y_a^{A,\ell m}\left(Y^a_{A',\ell' m'}\right)^*\mathrm{d}\Omega=N(A,r,\ell)\,\delta_{AA'}\delta_{\ell \ell'}\delta_{m m'},
\]
where \(\{A,A'\}\in\{E,B,R\}\) and \(N(A,r,\ell)\) is the specific normalization factor of the vector harmonic \(Y^a_{A,\ell m}\). Once a pure-spin vector harmonic basis is chosen, the natural progression is to also work with the pure-spin tensor harmonics from Thorne~\cite{Thorne}, here again with a different normalization:

\vspace{-2em}
\begin{multicols}{2}
\begin{align*}
T^{T0,\ell m}_{ab}&=\sigma_{ab}Y^{\ell m},\\
\\
T^{L0,\ell m}_{ab}&=n_an_bY^{\ell m},\\
\\
T^{E1,\ell m}_{ab}&=rn_{(a}\nabla_{b)}Y^{\ell m},
\end{align*}

\begin{align*}
T^{B1,\ell m}_{ab}&=rn^\wa_{(a}\epsilon_{b)c}^{\;\;\;\;\,d}n^c\,\nabla_{d}Y^{\ell m},\\
\\
T^{E2,\ell m}_{ab}&=r^2(\sigma_a^{\;c}\sigma_b^{\;d}-\frac12\sigma_{ab}\sigma^{cd})\nabla_c\nabla_dY^{\ell m},\\
\\
T^{B2,\ell m}_{ab}&=r^2\sigma_{(a}^{\;\;\,c}\epsilon_{b)e}^{\;\;\;\;\,d}n^e\,\nabla_c\nabla_dY^{\ell m}.
\end{align*}
\end{multicols}
As with the vector harmonics, these tensor harmonics are mutually orthogonal, 
\[
\oint T_{ab}^{A,\ell m}\left(T^{ab}_{A',\ell' m'}\right)^*\mathrm{d}\Omega=N(A,r,\ell)\,\delta_{AA'}\delta_{\ell \ell'}\delta_{m m'},
\]
where \(A\) and \(A'\) are understood to come from the set of tensor harmonics and \(N(A,r,\ell)\) is the associated tensor harmonic normalization of \(T^{ab}_{A,\ell m}\). These normalization functions \(N\) are detailed in Appendix~\ref{sec:a-kdecomp}, which outlines a generic decomposition of a tensor field \(h_{ab}\). A complete tensor basis may finally be constructed by symmetrizing \(v_a\) with the various vector harmonics, as seen in Eq.~\eqref{eq:hina-k}. 

The vector and tensor harmonics above that are purely tangent to the 2-sphere agree with the harmonics introduced by Martel and Poisson~\cite{MartelPoisson}. A similar basis was introduced for self-force calculations in the Lorenz gauge by Barack and Lousto \cite{BarackLousto}. We refer to Appendix A of Wardell and Warburton \cite{WardellWarburton} for an extensive summary of the Lorenz gauge conventions.
\section{A--K Notation}
\label{sec:a-knotation}

In perturbation theory, one writes the components of the physical metric \(g_{ab}\) as an expansion off of the background metric \(g_{ab}^0\) in terms of the tensor field given in Eq.~\eqref{hdef}, which to linear order takes the form,
\[
g_{ab}=g^0_{ab}+h_{ab}.
\]
Both the physical metric and the background metric are solutions to the Einstein Field Equations (EFEs), 
\begin{equation}
G_{ab}(g)=8\pi T_{ab},
\end{equation}
and we may expand the field equations in powers of the metric perturbation \(h_{ab}\),
\begin{equation}
\label{eq:EFEexpansion}
G_{ab}(g^0+h)=G_{ab}(g^0)-\frac{1}{2}E_{ab}(h)+O(h^2),
\end{equation}
with,
\begin{equation}
\label{eq:linearizedEFE}
E_{ab}(h)=\nabla^c\nabla_ch_{ab}+\nabla_a\nabla_bh^c_{\;c}-2\nabla_{(a}\nabla^ch_{b)c}+2R_{a\;\;b}^{\;\;c\;\;d}h_{cd}+g^0_{ab}(\nabla^c\nabla^dh_{cd}-\nabla^c\nabla_ch^d_{\;d}),
\end{equation}
where the covariant derivative operator is compatible with the background metric and \(R_{abcd}\) is the background curvature tensor. The operator \(E_{ab}\) is called the \textit{linearized Einstein operator}. We assume for this paper that the background metric is a solution to the vacuum EFEs (in particular, the Schwarzschild metric); as such, any perturbing matter present in the physical spacetime generates a stress-energy source for the linearized Einstein operator,
\begin{equation}
\label{eq:EFEstress}
E_{ab}(h)=-16\pi T_{ab},
\end{equation}
and one identifies that the tensor perturbation \(h_{ab}\) is the solution to this system of ten coupled, linear partial differential equations.

In his work, Detweiler introduced a convenient decomposition of the harmonic modes of the metric perturbation, utilizing a set of ten scalar functions and the tensor harmonic basis outlined in Sec.~\ref{sec:harmonicbasis}:
\begin{align}
\label{eq:hina-k}
\nonumber h_{ab}^{\ell m}&=\mathrm{A}\,v_av_bY^{\ell m}+2\,\mathrm{B}\,v^\wg_{(a}Y^{E,\ell m}_{b)}+2\,\mathrm{C}\,v^\wg_{(a}Y^{B,\ell m}_{b)}+2\,\mathrm{D}\,v^\wg_{(a}Y^{R,\ell m}_{b)}+\mathrm{E}\,T^{T0,\ell m}_{ab}+\mathrm{F}\,T^{E2,\ell m}_{ab}\\
&\;\;\;\;+\mathrm{G}\,T^{B2,\ell m}_{ab}+2\,\mathrm{H}\,T^{E1,\ell m}_{ab}+2\,\mathrm{J}\,T^{B1,\ell m}_{ab}+\mathrm{K}\,T^{L0,\ell m}_{ab}.
\end{align}
Here, all coefficients A through K are scalar functions of \((t,r)\), with the explicit functional dependence and harmonic indices suppressed for simplicity. We say that the metric perturbation above is written in ``A--K notation.'' Note that the the letter ``I'' is skipped in this decomposition to avoid confusion with the imaginary unit \(i\equiv\sqrt{-1}\).

Such a decomposition into scalar functions is certainly not new to the field of perturbation theory; one standard convention was introduced by Regge and Wheeler (RW) \cite{RW} (see comparisons below in Secs.~\ref{subsec:evenRWcompare} and \ref{subsec:oddRWcompare}). While the RW decomposition has become ubiquitous to the study of perturbations in the RW gauge (detailed in Sec.~\ref{subsec:RWgauge}), its notation choices can introduce unnecessary confusion when discussing certain components of the metric perturbation. For instance, the \(h_{t\theta}\) component of the metric perturbation is proportional to the scalar function \(h_0\) for \textit{both} even- and odd-parity, and is not to be confused with the scalar function \(H_0\), which represents the component \(h_{tt}\). While the notation may be clear in writing, confusion arises regularly in spoken discussion. The authors understand that while any choice of naming convention, such as the choice made by RW or the A--K notation above, is entirely arbitrary, Detweiler opted for clarity over compactness in choosing the A--K variables.

To ease those comfortable with the conventions of RW into the A--K notation, we find the correspondence between each A--K variable and its associated RW scalar function below. Writing the metric perturbation, Eq.~\eqref{eq:hina-k}, in a matrix form and expanding the tensor harmonics allows for a simple comparison with RW's conventions, their Eqs. (12)-(13).

\subsection{Even-Parity}
\label{subsec:evenRWcompare}

\noindent The even-parity pieces of the metric perturbation with \(\ell\geq2\) expand to:
\begin{equation*}
h^\text{even}_{ab}=\left(
\begin{tabular}{cccc}
\(\mathrm{A}\,Y_{\ell m}\)&\(-\mathrm{D}\,Y_{\ell m}\)&\(-r\mathrm{B}\,\partial_\theta Y_{\ell m}\)&\(-r\mathrm{B}\,\partial_\phi Y_{\ell m}\)\\
\\
\(-\mathrm{D}\,Y_{\ell m}\)&\(\mathrm{K}\,Y_{\ell m}\)&\(r\mathrm{H}\,\partial_\theta Y_{\ell m}\)&\(r\mathrm{H}\,\partial_\phi Y_{\ell m}\)\\
\\
\(\text{Sym}\)&\(\text{Sym}\)&\(r^2\left[\mathrm{E}+\mathrm{F}\left(\partial_\theta^2+\frac12\ell(\ell+1)\right)\right]Y_{\ell m}\)&\(\text{Sym}\)\\
\\
\(\text{Sym}\)&\(\text{Sym}\)&\(r^2\mathrm{F}\left[\partial_\theta\partial_\phi -\cot\theta\partial_\phi\right]Y_{\ell m}\)&\makecell{\(r^2\sin^\theta\left[\mathrm{E}-
 \mathrm{F}(\partial^2_\theta
+\frac12\ell(\ell+1))\right]Y_{\ell m}\)}
\end{tabular}
\right).
\end{equation*}
After comparison with Eq. (13) of Regge and Wheeler, the connection between the RW variables and A--K is apparent:
\begin{align}
\label{eq:RWtoAKeven}
\nonumber\mathrm{A}&=\left(1-\frac{2M}{r}\right)H_0^\text{rw},\\
\nonumber\mathrm{B}&=-\frac1rh_0^\text{rw, ev},\\
\nonumber\mathrm{D}&=-H_1^\text{rw},\\
\mathrm{E}&=K^\text{rw}-\frac12\ell(\ell+1)G^\text{rw},\\
\nonumber\mathrm{F}&=G^\text{rw},\\
\nonumber\mathrm{H}&=\frac1rh_1^\text{rw, ev},\\
\nonumber\mathrm{K}&=\left(1-\frac{2M}{r}\right)^{-1}H_2^\text{rw}.
\end{align}
Notice the difference between Detweiler's E and the RW \(K^\text{rw}\), even though both terms are associated with the two-sphere metric \(\sigma_{ab}\). This change results from RW's use of a different even-parity tensor harmonic on the two-sphere, \(\sigma_a^{\;c}\sigma_b^{\;d}\nabla_c\nabla_cY_{\ell m}\), instead of the trace-free counterpart \(T^{E2}_{ab}\) used here.

One might be worried about the apparent disorder of the listing in Eqs.~\eqref{eq:RWtoAKeven}. Why are A, B, and D present, yet C is absent? This is discussed in App.~\ref{sec:a-kdecomp}.

\subsection{Odd-Parity}

\label{subsec:oddRWcompare}

\noindent We also perform the corresponding comparison for odd-parity, again with \(\ell\geq2\):
\begin{equation*}
h^\text{odd}_{ab}=\left(
\begin{tabular}{cccc}
0&0&\(\frac{r}{\sin\theta}\mathrm{C}\,\partial_\phi Y_{\ell m}\)&\(-r\sin\theta\,\mathrm{C}\,\partial_\theta Y_{\ell m}\)\\
\\
0&0&\(-\frac{r}{\sin\theta}\mathrm{J}\,\partial_\phi Y_{\ell m}\)&\(r\sin\theta\,\mathrm{J}\,\partial_\theta Y_{\ell m}\)\\
\\
Sym&Sym&\(-\frac{r^2}{\sin\theta}\,\mathrm{G}[\partial_\theta\partial_\phi-\cot\theta\,\partial_\phi]Y_{\ell m}\)&Sym\\
\\
Sym&Sym&\makecell{\(-\frac{r^2}{2}\,\mathrm{G}[\frac1{\sin\theta}\partial^2_\phi+\cos\theta\,\partial_\theta\)\\ \(-\sin\theta\,\partial^2_\theta]Y_{\ell m}\)}&\(r^2\mathrm{G}[\sin\theta\,\partial_\theta\partial_\phi-\cos\theta\,\partial_\phi]Y_{\ell m}\)\\
\end{tabular}
\right)_.
\end{equation*}
Looking at Eq. (12) of RW, we may ``pick off'' the relationship between each RW variable and the A--K variables in the odd-parity sector:
\begin{align}
\label{eq:RWtoAKodd}
\nonumber\mathrm{C}&=- r^{-1}h_0^\text{rw, od},\\
\mathrm{G}&=- r^{-2}h_2^\text{rw},\\
\nonumber\mathrm{J}&=r^{-1}h_1^\text{rw, od}.
\end{align}
The correspondence between the A--K variables and the RW metric components allows anyone comfortable with the analysis of Regge and Wheeler to immediately use the same basic intuitions when dealing with the A--K variables.

\subsection{A--K Projections of Generic Tensors}
\label{sec:a-kproj}
The A--K variables in Eq.~\eqref{eq:hina-k} are defined by projecting the metric perturbation onto the tensor harmonic basis chosen in Sec.~\ref{sec:harmonicbasis}. The prescription may be expanded to describe the projections of any rank 2 tensor in the Schwarzschild background. 

For example, it may be useful to look at the A--K components of a generic tensor \(X_{ab}\). From App.~\ref{sec:a-kdecomp}, we may write e.g.,
\begin{equation}
\label{eq:TA}
X_\text{K}=\left(1-\frac{2M}{r}\right)^{-2}\oint T_{L0}^{ab*}X_{ab}\,\mathrm{d}\Omega.
\end{equation}
We call \(X_\tK\) the ``K term'' of the tensor \(X_{ab}\). The linearized EFEs from Eq.~\eqref{eq:linearizedEFE} may also be written as A--K component equations such as,
\begin{equation}
\label{eq:linearizedproj}
E_\tK=-16\pi T_\tK,
\end{equation}
where the right-hand-side of Eq.~\eqref{eq:linearizedproj} is the K term of the stress-energy tensor \(T_{ab}\), and the left-hand-side involves various combinations of the A--K variables of the metric perturbation and their \(t\)- and \(r\)-derivatives, as listed in App.~\ref{sec:einsteina-k}.

\section{Gauge Transformations in A--K Notation}
\label{sec:gaugeina-k}
When a gauge transformation is performed using a gauge vector \(\xi^a\), the first-order metric perturbation \(h_{ab}\) is changed by  the symmetrized covariant derivative of the gauge vector, \(2\nabla_{(a}\xi_{b)}\), as seen in Eq.~\eqref{gaugetrans4},
\begin{equation}
\label{eq:gaugetransformation}
h^\text{new}_{ab}=h^\text{old}_{ab}-2\nabla_{(a}\xi_{b)}.
\end{equation}
To better understand the effects of such a transformation \eqref{eq:gaugetransformation} on the A--K projections of the metric perturbation, we decompose the gauge vector into the pure-spin harmonic basis adopted in Sec.~\ref{sec:harmonicbasis},
\begin{equation}
\label{eq:gaugevecdecomp}
\xi_a=\tP\,v_aY_{\ell m}+\tR\,n_aY_{\ell m}+\tS\,Y^{E,\ell m}_{a}+\tQ\,Y^{B,\ell m}_{a}.
\end{equation}
The letters P, R, S, and Q represent four scalar functions of \((t,r)\) with harmonic labels and coordinate dependence suppressed for convenience. The functions P, R, and S describe the three degrees of gauge freedom present in the even-parity sector, and the function Q contains the one degree of odd-parity gauge freedom available. We next find the A--K projections of \(2\nabla_{(a}\xi_{b)}\) (see Sec.~\ref{sec:a-kproj}), writing e.g.,
\begin{align}
 \Delta\text{A}&\equiv2\left(1-\frac{2M}{r}\right)^2\oint v^av^b\nabla_{a}\xi_{b}Y^*_{\ell m}\,\mathrm{d}\Omega,
 \end{align}
 for the A term of the change on the metric perturbation induced by the gauge vector. This term \(\Delta \text{A}\) alone is responsible for changes to the A term of \(h_{ab}\),
 \begin{equation}
 \tA_\text{new}=\tA_\text{old}-\Delta \tA.
 \end{equation} 
 The ``new'' and ``old'' subscripts correspond to projections of \(h^\text{new}_{ab}\) and \(h^\text{old}_{ab}\), respectively. Once all ten projections of \(2\nabla_{(a}\xi_{b)}\)  have been computed, applying the gauge transformation becomes a matter of subtraction. The A--K components of \(2\nabla_{(a}\xi_{b)}\) are found below.
\vfill\eject
\newpage
\begin{multicols}{2}
\begin{align*}
\nonumber\Delta\text{A}&=-2\tder\tP-\frac{2M(r-2M)}{r^3}\tR,\\
\nonumber\\
\nonumber\Delta\text{B}&=\frac1r\tP-\tder \tS,\\
\nonumber\\
\nonumber\Delta\text{D}&=\left(\rd-\frac{2M}{r(r-2M)}\right)\tP-\tder \tR,\\
\nonumber\\
\nonumber\Delta\text{E}&=\frac{2(r-2M)}{r^2}\tR-\frac{\ell(\ell+1)}{r}\tS,\\
\nonumber\\
\nonumber\Delta\text{F}&=\frac{2}{r}\tS,\\
\nonumber\\
\nonumber\Delta\text{H}&=\frac1r\tR+\left(\rd-\frac1r\right)\tS,\\
\nonumber\\
\nonumber\Delta\text{K}&=\left(2\rd+\frac{2M}{r(r-2M)}\right)\tR,
\end{align*}

\begin{align}
\label{eq:AKtoPQRS}
\nonumber\\
\nonumber\\
\nonumber\\
\nonumber\\
\nonumber\\
\nonumber\\
\nonumber\Delta\text{C}&=-\tder \tQ,\\
\nonumber\\
\Delta\text{G}&=\frac2r\tQ,\\
\nonumber\\
\nonumber\Delta\text{J}&=\left(\rd -\frac1r\right)\tQ.
\end{align}
\end{multicols}
\noindent We have listed the changes to the even-parity components of the metric on the left and to the odd-parity components on the right. The beauty of this decomposition is revealed in the ease with which one may use Eqs.~\eqref{eq:AKtoPQRS} to choose a privileged gauge in a Schwarzschild background; two very obvious choices for the components of the gauge vector lead to both the RW and EZ gauges, which we now discuss.

\subsection{Regge-Wheeler Gauge}
\label{subsec:RWgauge}
The RW gauge is defined by setting certain RW variables to zero, as was first done by Regge and Wheeler~\cite{RW}. In the even-parity sector, with three degrees of gauge freedom available, RW choose to eliminate,
\[
h_0^\text{rw, ev}=h_1^\text{rw, ev}=G^\text{rw}=0.
\]
Now that we have an association between the RW variables and A--K, the RW gauge choice in A--K notation follows from the mapping listed in Sec.~\ref{sec:a-knotation}. The even-parity choice of RW gauge corresponds to setting,
\[
\tB=\tF=\tH=0,
\]
and also brings \(\tE=K^\text{rw}\). Using these gauge conditions and Eqs.~\eqref{eq:AKtoPQRS}, it is a straightforward process to recover the associated even-parity components of the gauge vector.

For an example, we analyze the procedure for the F term. Starting with a metric perturbation in an arbitrary gauge, we calculate \(\tF_\text{old}\).  From Eqs.~\eqref{eq:AKtoPQRS}, the equation governing the effect of the gauge vector on \(\tF_\text{old}\) is,
\begin{equation}\label{eq:Fchange}
\Delta F = \frac2r \tS.
\end{equation}
As our gauge condition eliminates \(\tF_\text{new}\) in the RW gauge, we are left to solve the algebraic equation
\[
0=\tF_\text{old}-\frac2r\tS,
\]
resulting in \(\tS=r\tF_\text{old}/2\). We have now fixed this degree of gauge freedom, and substitute the solution for S back into Eqs.~\eqref{eq:AKtoPQRS}. We then move on to the next component to be eliminated, and repeat the procedure. The result of this transformation to the RW gauge in the even-parity sector gives us the following choices for P, R, and S:
\begin{align*}
\tP&=r\tB_\text{old}+r\tder\left(\frac r2\tF_\text{old}\right) \\
\tR&=r\tH_\text{old}-\left(\rd-\frac 1r\right)\frac r2\tF_\text{old}\\
\tS&=\frac r2\tF_\text{old}.
\end{align*}

In the odd-parity sector there is one degree of gauge freedom, and RW choose to eliminate
\(
h_2^\text{rw}=0,
\)
thus setting,
\[
\tG=0,
\]
in the A--K notation. This amounts to choosing \(\tQ=\frac r2 \tG_\text{old}\) for the gauge vector. In summary, the RW gauge condition in A--K notation is written,
\[
\tB=\tF=\tH=\tG=0,
\]
and the transition to the RW gauge from an arbitrary gauge is accomplished through the gauge vector,
\begin{align}
\label{eq:gaugevecRW}
\xi_a^{rw}&=\left(r\text{B}_\text{old}+\frac{r^2}{2}\frac{\partial\text{F}_\text{old}}{\partial t}\right)\,v_aY_{\ell m}+\left(r\text{H}_\text{old}-\frac r2\frac{\partial\text{F}_\text{old}}{\partial r}\right)n_aY_{\ell m}\\
&\,\,\,\,\,\,+\frac r2\text{F}_\text{old}\,Y^{E,\ell m}_{a}+\frac r2 \text{G}_\text{old}\,Y^{B,\ell m}_{a}.\nonumber
\end{align}

It is an important fact to notice that each component of the gauge vector \eqref{eq:gaugevecRW} has been \textit{uniquely} determined, without the introduction of arbitrary functions of \(t\) or \(r\). This arises from the fact that the RW gauge uses the entirety of the allowed gauge freedom~\cite{Nakano}, and is discussed in further detail in Sec.~\ref{subsec:othergauges}.

\subsection{Easy Gauge}
\label{subsec:EZgauge}
The EZ gauge used by Detweiler differs from the RW gauge, in that the metric components chosen to vanish are,
\[
\mathrm{B}=\mathrm{E}=\mathrm{F}=\mathrm{G}=0.
\]
In choosing the above components to vanish and following the same procedure outlined in Sec.~\ref{subsec:RWgauge}, the gauge vector components become,
\begin{align*}
\text{P}&=r\text{B}_\text{old}+r\frac{\partial}{\partial t}\left(\frac r2\text{F}_\text{old}\right) \\
\text{R}&=\frac{r^2}{2(r-2M)}\left(\text{E}_\text{old}+\frac{\ell(\ell+1)}{2}\text{F}_\text{old}\right)\\
\text{S}&=\frac r2\text{F}_\text{old}\\
\text{Q}&=\frac r2 \text{G}_\text{old},
\end{align*}
yielding a gauge vector,
\begin{align*}
\xi_a^{ez}&=\left(r\text{B}_\text{old}+\frac{r^2}{2}\frac{\partial\text{F}_\text{old}}{\partial t}\right)\,v_aY_{\ell m}+\left(\frac{r^2}{2(r-2M)}\left[\text{E}_\text{old}+\frac{\ell(\ell+1)}{2}\text{F}_\text{old}\right]\right)n_aY_{\ell m}\\
&\,\,\,\,\,\,+\frac r2\text{F}_\text{old}\,Y^{E,\ell m}_{a}+\frac r2 \text{G}_\text{old}\,Y^{B,\ell m}_{a}.
\end{align*}
We see that the difference between the two gauge vectors for the RW and EZ gauges is purely longitudinal. Again, all degrees of gauge freedom have been used, so the EZ gauge is entirely determined, and we see that the \textit{entire} metric perturbation has been set to vanish on the 2-sphere. Finally we note that the translation between the EZ gauge metric components in A--K notation and the RW notation listed in Eqs.~\eqref{eq:RWtoAKeven} is quite straightforward, as the F term is set to zero (just as in the RW gauge). This eliminates the mixing between \(K^\text{rw}\) and \(G^\text{rw}\) in the even-parity sector when solving for E; we are left with \(\mathrm{E}=K^\text{rw}\) exactly, and we may then simply write the EZ gauge condition in the RW variables as,
\[
h_0^\text{rw, ev}=G^\text{rw}=K^\text{rw}=h_2^\text{rw}=0.
\]

\subsection{Other Gauges}
\label{subsec:othergauges}
Finally, we comment on how other useful gauge choices might be made through the prescription outlined in Sec.~\ref{subsec:RWgauge}. It can be seen in Eqs.~\eqref{eq:AKtoPQRS} that one odd-parity component and five even-parity components of the metric are \textit{algebraically} related to the components of the gauge vector in this decomposition; for instance, the change in the G term of the metric perturbation under a gauge transformation is directly proportional to the Q component of the gauge vector. Similar relationships hold for certain even-parity components of the metric as well, e.g., the change in the F term is proportional to S.

Choosing to set the component \(\tG=0\) requires us to specify the function Q, but this is done entirely algebraically and without integration. One could choose the function Q to eliminate C or J instead, but this would be done at the cost of introducing arbitrary functions of \(r\) or \(t\), respectively, via the integration necessary to determine Q. Such functions would contain unused gauge information and it would be necessary to fix these functions before the gauge choice is unique.

A similar situation exists in the even-parity sector, except we have more degrees of gauge freedom that must be specified. The simplest way to choose a unique gauge is by first setting \(\tF=0\), which fixes the function S. Once S is chosen, we may uniquely determine P by setting \(\tB=0\) as well. With those two gauge vector components determined, we are left with three remaining pieces of the metric which are algebraically related to our one remaining unspecified degree of gauge freedom found in R, specifically \(\Delta \tH\), \(\Delta \tE\), and \(\Delta \tA\). Using R to set \(\tH=0\) is the RW gauge choice, and using R to set \(\tE=0\) is the EZ gauge choice.



The choice \(\tA=0\) introduces a ``\(tt\)-free, radiation-free'' gauge, in which the A, B, and F terms of the even-parity decomposition are set to zero, along with the odd-parity G. This unique gauge choice eliminates \(h_{tt}\) and all spin-2 modes of the metric perturbation, along with the piece of the metric proportional to \(v_{(a}Y^{E}_{b)}\). Whether or not such a gauge is useful would depend on the problem being solved.

\section{Gauge Invariants}
\label{sec:gaugeinva-k}

The goal of this section is to find a general set of gauge invariant quantities in the Schwarzschild spacetime. Gauge invariant quantities are vital to the study of motion in General Relativity as they allow for the comparison of results calculated using different working coordinate systems, in accordance to the general principle of covariance.

The approach used to find gauge invariants in the Schwarzschild spacetime below follows in a similar manner to that of Gerlach and Sengupta (GS) \cite{Gerlach}. It is possible to transform the gauge invariant quantities found in Eqs.~\eqref{eq:indepgaugeinvar} into the gauge invariants listed by GS in their Eqs.~(7a) and (8a). Further discussion of the GS decomposition and gauge choices may be found in Brizuela \textit{et al.} \cite{brizuela}.

\subsection{General Approach}
\label{sec:generalapproach}
 Having in hand the effects of a gauge transformation on the metric projections in Eqs.~\eqref{eq:AKtoPQRS}, e.g., \(\mathrm{J}\rightarrow \mathrm{J}-\Delta\mathrm{J}\), we follow the comments made in Sec.~\ref{sec:gaugeinvintro} to find the gauge invariants. We note that Eqs.~\eqref{eq:AKtoPQRS} can be inverted to find the components of \(\xi^a\) and their derivatives in terms of changes in the metric under the gauge transformation. For example, we find
 \begin{align}
\text{Q}&=\frac{1}{2}r\Delta \text{G},  \label{Q:1} \\
\frac{d\text{Q}}{dr} &= \Delta \text{J} + \frac{1}{2}\Delta \text{G} \label{Q:2}.
 \end{align}
 Taking the derivative of Eq.~\eqref{Q:1} with respect to \(r\) and subtracting Eq.~\eqref{Q:2} gives 
 \begin{align}
	\Delta \text{J}-\frac r2\ed{r}{\(\Delta\) G}=0. \label{alphahector}
 \end{align}
 Thus, 
 \begin{align}
\alpha&=\text{J}-\frac r2\ed{r}{G} \label{alphahector2},
\end{align}
is a gauge invariant quantity, as the action of the gauge vector on J and G will exactly cancel in this combination. This simple, worked example outlines the general approach taken to finding the gauge invariants in the A--K notation, which will be applied both for the general \(\ell\geq2\) case below and the special cases of \(\ell=0,~1\) in the sections to follow.

\subsection{General Gauge Invariants for \(\ell\geq2\)}
\label{sec:lg2invariants}
Repeating the procedure of Sec.~\ref{sec:generalapproach} for all components and derivatives of \(\xi^a\) yields the following six independent gauge invariant quantities, two of odd-parity and four of even-parity,
\begin{align}
\label{eq:indepgaugeinvar}
\nonumber\alpha&=\text{J}-\frac r2\ed{r}{G},\\
\nonumber\\
\nonumber\beta&=-\text{C}-\frac r2\ed{t}{G},\\
\nonumber\\
\chi&=\text{H}-\frac{r}{2(r-2M)}\text{E}-\frac{\ell(\ell+1)r}{4(r-2M)}\text{F}-\frac r2\ed{r}{F},\\
\nonumber\\
\nonumber\psi&=\frac12\text{K}-\frac{r(r-3M)}{2(r-2M)^2}\text{E}-\frac{r^2}{2(r-2M)}\ed{r}{E}-\frac{\ell(\ell+1)r(r-3M)}{4(r-2M)^2}\text{F}-\frac{\ell(\ell+1)r^2}{4(r-2M)}\ed{r}{F},\\
\nonumber\\
\nonumber\delta&=\text{D}+\frac{r^2}{2(r-2M)}\ed{t}{E}-\frac{r-4M}{r-2M}\text{B}-r\ed{r}{B}-\frac{r^2}2\edm{t}{r}{F}+\frac{r\left[4(3M-r)+r\ell(\ell+1)\right]}{4(r-2M)}\ed{t}{F},\\
\nonumber\\
\nonumber\epsilon&=-\frac12\text{A}-\frac{M}{2r}\text{E}-r\ed{t}{B}-\frac{\ell(\ell+1)M}{4r}\text{F}-\frac{r^2}2\edt{t}{F}.
\end{align}
Further gauge invariants may be found by combining various derivatives of these six gauge invariants; two examples of these constructed gauge invariant quantities are given, one of each parity,
\begin{align}
\label{eq:gammaGI}\gamma&=\tder\alpha-\rd\beta\\
\nonumber&=\ed{r}{C}+\ed{t}{J}+\frac12\ed{t}{G},\\
\nonumber\\
\label{eq:omGI}o&=\tder\delta-\rd\epsilon
\\
\nonumber&=\frac12\ed{r}{A}+\frac{2M}{r-2M}\ed{t}{B}+\frac{\partial}{\partial t}\text{D}+\frac{r^2}{2(r-2M)}\edt{t}{E}+\frac{r[4M+r\ell(\ell+1)]}{4(r-2M)}\edt{t}{F}\\
\nonumber&\;\;\;\;\;\;+\frac{M}{2r}\left(\frac{\partial}{\partial r}-\frac1r\right)\text{E}+\frac{\ell(\ell+1)M}{4r}\left(\frac{\partial}{\partial r}-\frac1r\right)\text{F}.
\end{align}

\subsection{Linearized Einstein Field Equations}
\label{subsec:LEFE}

The projections of the linearized EFEs introduced in Sec.~\ref{sec:a-kproj} play an important role in the study of gauge invariants, as they themselves are manifestly gauge invariant quantities at first order. It is to be expected, then, that each A--K projection of the linearized EFEs may be written fully as a combination of the gauge invariants in Eqs.~\eqref{eq:indepgaugeinvar}. This is done in App.~\ref{sec:einsteina-kgauge}, and will prove useful in the remainder of this paper when searching for equations governing the gauge invariants themselves.

\section{Metric Reconstruction at \(\ell\geq2\)}\label{sec:metricrecon}
With the gauge invariants calculated from Eqs.~\eqref{eq:indepgaugeinvar}, we investigate how to recover the metric components from these gauge invariant quantities. We now assume that a given perturbing stress-energy, \(T_{ab}\), is known. From Eq.~\eqref{eq:EFEstress} and the example in Sec.~\ref{sec:a-kproj}, we interpret the projections of the linearized EFEs in App.~\ref{sec:einsteina-kgauge} as source terms for the coupled differential equations governing the gauge invariants. Finding solutions to these equations will then allow us to determine the metric components by inverting Eqs.~\eqref{eq:indepgaugeinvar}, after choosing the gauge of our liking. Should one wish to perform vacuum perturbations, one simply sets e.g., \(E_\tK=0\), in App.~\ref{sec:einsteina-kgauge} for all projections of the EFEs.


\subsection{Even-Parity}
It is clear from the Bianchi identities that not all seven even EFE projections are independent. We find a proper set of equations to be \(E_\text{A}\), \(E_\text{D}\), \(E_\text{F}\), and \((E_\text{K}+2E_\text{H})\). These four equations allow us to construct a system of coupled partial differential equations for \(\chi\) and \(\psi\) in terms of stress-energy projections:
\begin{align}
\label{eq:PDEchi}\frac{\partial}{\partial r}\chi&=\frac{1}{(6M+r\lambda)}
\left[
-\frac{4r^2}{\ell(\ell+1)}\frac{\partial^2}{\partial t^2}\psi
+\frac{(4M+r\lambda)}{r}\psi
+\frac{2r^3}{(r-2M)}\frac{\partial^2}{\partial t^2}\chi
-\frac{M(\ell^2+\ell+2)}{r-2M}\chi\right.\\
\nonumber&\left.\qquad +\frac{r^4}{(r-2M)\ell(\ell+1)}\frac{\partial}{\partial t}E_\text{D}-\frac{r^2(6M+r\lambda)}{2(r-2M)}E_\text{F}
+r^2E_\text{H}+\frac{r^2}{2}E_\text{K}\right],\\
\nonumber\\
\frac{\partial}{\partial r}\psi&=\frac{1}{(6M+r\lambda)}
\left[
-\frac{2r^3}{r-2M}\frac{\partial^2}{\partial t^2}\psi
-\frac{12M^2+2rM-2r^2+(3rM+r^2)\ell(\ell+1)}{r(r-2M)}\psi\right.\label{eq:PDEpsi}\\
\nonumber&\left.\qquad+\frac{r^4\ell(\ell+1)}{(r-2M)^2}\frac{\partial^2}{\partial t^2}\chi
-\frac{\ell(\ell+1)[(2rM-r^2)\ell(\ell+1)+9M^2+2r^2-8rM]}{(r-2M)^2}\chi
-\frac{r^4(6M+r\lambda)}{4(r-2M)^3}E_\text{A}\right.\\
\nonumber&\left.\qquad +\frac{r^5}{2(r-2M)^2}\frac{\partial}{\partial t}E_\text{D}-\frac{r^3\ell(\ell+1)(6M+r\lambda)}{4(r-2M)^2}E_\text{F}
+\frac{r^3\ell(\ell+1)}{2(r-2M)}E_\text{H}
+\frac{r^3\ell(\ell+1)}{4(r-2M)}E_\text{K}
\right],
\end{align}
with \(\lambda=(\ell-1)(\ell+2)\). Note that our definition of \(\lambda\) here differs from some other choices found in the literature, in that it lacks a factor of \(1/2\). While these equations are certainly not simple, they have one special feature in that they do not involve radial derivatives of the source terms. In applications where the perturbing stress-energy is singular in some region of space (such as a delta function source), eliminating spatial derivatives of the source terms may improve the efforts of regularization\cite{Vega}. 

We find the two remaining even-parity gauge invariants \(\delta\) and \(\epsilon\) from \(E_\text{D}\) and \(E_\text{F}\), respectively, written in terms of \(\chi\) and \(\psi\):
\begin{align}
\label{eq:deltaGIsol}
\delta&=\frac{r^2}{\ell(\ell+1)}\left[-E_\text{D}+\frac{4(r-2M)}{r^2}\frac{\partial}{\partial t}\psi-\frac{\ell(\ell+1)}{r}\frac{\partial}{\partial t}\chi\right],\\
\nonumber\\
\label{eq:epsGIsol}\epsilon&=\frac{r(r-2M)}{2}\left[E_\text{F}-\frac{2(r-2M)}{r^3}\psi+\frac{2}{r^2}\chi+\frac{2(r-2M)}{r^2}\frac{\partial}{\partial r}\chi\right].
\end{align}
The equations for \(\delta\) and \(\epsilon\) also do not involve spatial derivatives of source terms; even the spatial derivative of \(\chi\) on the right-hand side of Eq.~\eqref{eq:epsGIsol} may be removed with Eq.~\eqref{eq:PDEchi}. However, the direct proportionality between e.g. \(\delta\) and \(E_\tD\) shows the possible singular behavior of the metric components in this formalism, a well-known phenomenon which arises in metric reconstruction from a delta function source \cite{HopperEvans}.


Our final step before finding the metric components is to choose a particular gauge in which to work. We will adopt the EZ gauge, setting
\[
\text{B}=\text{E}=\text{F}=0.
\]
Once this particular gauge is chosen in Eqs.~\eqref{eq:indepgaugeinvar}, the four independent gauge invariants reduce to the simple expressions,
\begin{align}
\label{eq:indepgaugeinvarEZgauge}
\nonumber\chi&=\text{H},\\
\psi&=\frac12\text{K},\\
\nonumber\delta&=\text{D},\\
\nonumber\epsilon&=-\frac12\text{A},
\end{align}
which are trivial to invert for the metric components A, K, D, and H after having integrated Eqs.~\eqref{eq:PDEchi}-\eqref{eq:PDEpsi}. Should one wish to find the metric components in the RW gauge (\(\tB=\tF=\tH=0\)) instead, one arrives at the following expressions for the metric components,
\begin{align}
\label{eq:indepgaugeinvarRWgauge}
\nonumber\tA&=-2\epsilon+\frac{2M(r-2M)}{r^2}\chi\\
\tD&=\delta+r\frac{\partial}{\partial t}\chi,\\
\nonumber\tE&=-\frac{2(r-2M)}{r}\chi\\
\nonumber\tK&=2\psi-\frac{2(r-M)}{r-2M}\chi-2r\frac{\partial}{\partial r}\chi.
\end{align}

An equally valid approach to metric reconstruction would be to investigate this problem in the vein of Zerilli \cite{zerilli}. Combining the chosen set of EFEs listed at the beginning of this section, one finds that the gauge invariants \(\chi\) and \(\psi\) may be combined into a single \textit{master variable} named after Zerilli,
\begin{align}
\label{eq:zerilli}
Z &= \frac{r-2M}{6M+\lambda r}\left[2(r-2M)\psi-\ell(\ell+1)r\chi \right].
\end{align}
This master variable obeys a sourced wave equation known as the Zerilli equation,
\begin{equation}
\label{eq:zerillieq}
\left(\frac{r}{r-2M}\frac{\partial^2}{\partial t^2}-\frac{r-2M}{r}\frac{\partial^2}{\partial r^2}-\frac{2M}{r^2}\frac{\partial}{\partial r}+\frac{\lambda^2(\lambda+2)r^3+6M(\lambda^2 r^2+6M\lambda r + 12M^2)}{r^3(6M+\lambda r)^2}\right)Z=S_\text{Z}.
\end{equation}
We list the (rather lengthy) source terms, \(S_\text{Z}\), at the end of App.~\ref{sec:einsteina-kgauge} in Eq.~\eqref{eq:zerillieqsource}. Our source differs from that of the Zerilli-Moncrief master variable \cite{MartelPoisson}. Using Eq.~\eqref{eq:zerillieq} grants us the added benefit of solving a hyperbolic wave equation, instead of a system of coupled equations, at the cost of introducing radial derivatives to the source terms. Depending on the problem being solved, this may or may not be an issue. Once one has integrated the Zerilli equation, metric reconstruction follows through use of Eqs.~\eqref{eq:zerilli} and either of Eqs.~\eqref{eq:PDEchi} or \eqref{eq:PDEpsi} to recover \(\chi\) and \(\psi\),
\begin{align}
\label{eq:chifromZ}
\chi&=\frac{-1}{(\lambda+2)(6M+\lambda r)}\left(2(6M+\lambda r)\frac{\partial}{\partial r}Z+\frac{\lambda(\lambda+2)r^2+6M(r\lambda + 4M)}{r(r-2M)}Z+\frac{r^4}{r-2M}E_\tA\right),\\
\nonumber\\
\label{eq:psifromZ}\psi&=\frac{-1}{(r-2M)^2(6M+\lambda r)}\left(r(r-2M)(6M+\lambda r)\frac{\partial}{\partial r}Z+(r^2\lambda - 3rM\lambda - 6M^2)Z+\frac{r^5}{2}E_\tA\right),
\end{align}
and then continues as above with Eqs.~\eqref{eq:deltaGIsol} and \eqref{eq:epsGIsol}.

The history of black hole perturbations is heavily entwined with the history of the Zerilli equation (and the Regge-Wheeler equation in the following section), and the literature is rich with discussions on particular solutions to these equations. The review by Nagar and Rezzolla \cite{NagarRezzolla} is an excellent place to start for anyone interested in reading about these various conventions. 

\subsection{Odd-Parity}
The contracted Bianchi identities tell us that only two of the odd-parity EFEs in App.~\ref{sec:a-kgaugeodd} are independent, corresponding to two equations for the invariants \(\alpha\) and \(\beta\). One may wish to find a set of coupled equations for these gauge invariants, as was done in the even-parity case, and also eliminate any radial derivatives of the source terms, but it is unclear how to find a combination of the EFEs that evolve well with Cauchy initial data under those conditions. Instead, to handle the odd-parity sector we introduce an odd-parity master variable equivalent to the Cunningham-Price-Moncrief function \cite{CPM} (their variable \(\tilde{\pi}_1\)),
\begin{align}
\label{eq:Wdef}
\frac{1}{r^2}W=\frac{\partial}{\partial t}\alpha - \frac{\partial}{\partial r}\beta + \frac{1}{r}\beta.
\end{align}
We find from the equations for \(E_\text{C}\) and \(E_\text{J}\) that \(W\) must satisfy a version of the Regge-Wheeler equation \cite{RW} with different source terms,
\begin{align}
\label{eq:RWeq}
\left(-\frac{r}{r-2M}\frac{\partial^2}{\partial t^2}+\frac{r-2M}{r}\frac{\partial^2}{\partial r^2}+\right.&\left.\frac{2M}{r^2}\frac{\partial}{\partial r}-\frac{r\ell(\ell+1)-6M}{r^3}\right)W\\
\nonumber&\qquad=r^2\frac{\partial}{\partial r}E_\text{C}+rE_\text{C}+r^2\frac{\partial}{\partial t}E_\text{J},
\end{align}
and we may re-express the gauge invariants in terms of \(W\) and source terms:
\begin{align}
\label{eq:alpha}\alpha&=-\frac{r}{\lambda}\left(\frac{1}{(r-2M)}\frac{\partial}{\partial t}W+rE_\text{J}\right),\\
\nonumber\\
\label{eq:beta}\beta&=-\frac{1}{\lambda}\left(\frac{r-2M}{r}\frac{\partial}{\partial r}W+\frac{r-2M}{r^2}W-r^2E_\text{C}\right).
\end{align}
It is immediately clear from Eq.~\eqref{eq:alpha} and Eq.~(5.18) of Martel and Poisson \cite{MartelPoisson} that the gauge invariant \(\alpha\) used in this paper is, in fact, directly related to the master variable introduced by RW in their original paper, \(\Phi\), with a different normalization,
\begin{align}
\Phi &=\frac{r-2M}{r}\alpha.
\end{align}
 As in the even-parity case, we now choose to adopt the EZ gauge condition \(\text{G}=0\) (which is identical to the RW gauge condition) and solve for the two remaining metric components,
\begin{align}
\alpha&=\text{J},\\
\beta&=-\text{C},
\end{align}
which fully recovers the metric perturbation.


\section{Gauge Invariants for \(\ell=0,1\)}\label{GI:l=0,1}
The general approach to finding the gauge invariants of Sec.~\ref{sec:gaugeinva-k} may also be utilized for the specific cases of \(\ell=0,1\), despite the reduced number of degrees of freedom in each case. Below, we outline the structure of these gauge invariants, along with the metric reconstruction techniques used to recover the metric perturbation at the values of \(\ell=0,1\).

\subsection{Special Case: \(\ell=0\)}
\label{sec:ellzero}
For the special case of \(\ell=0\),  Eq.~\eqref{eq:hina-k} takes on a simple form,
\begin{align}
\label{eq:ellzeroh}
h_{ab}^{00}&=\frac{1}{2\sqrt{\pi}}\left(\text{A}_0\,v_av_b+2\text{D}_0\,v_{(a}n_{b)}+\text{E}_0\,\sigma_{ab}+\text{K}_0\,n_an_b\right),
\end{align}
with a zero subscript reminding us that we have set \(\ell=0\). The expression for the gauge vector reduces to give two explicit degrees of gauge freedom,
\begin{align}
\label{eq:gaugelzero}
\xi_a^0=\frac{1}{2\sqrt{\pi}}\left(\text{P}_0\,v_a+\text{R}_0\,n_a\right).
\end{align}
Looking to Eqs.~\eqref{eq:AKtoPQRS}, we construct the following two gauge invariant quantities in a similar fashion to those of Sec.~\ref{sec:generalapproach},
\begin{align}
\label{eq:gaugeinvlzero}
\nonumber\psi_0&=\frac12\tK_0-\frac{r(r-3M)}{2(r-2M)^2}\tE_0-\frac{r^2}{2(r-2M)}\rd\tE_0,\\
\nonumber\\
o_0&=\frac12\frac{\partial}{\partial r}\text{A}_0-\frac{M}{r(r-2M)}\text{A}_0+\tder\text{D}_0+\frac{r^2}{2(r-2M)}\tdd\text{E}_0\\
\nonumber&\,\,\,\,\,\,-\frac{M}{2r(r-2M)}\text{E}_0+\frac{M}{2r}\rd\text{E}_0.
\end{align}
It should be noted here that \(o_0\) is \textit{not} an \(\ell=0\) reduction of the gauge-invariant with the same name listed in Eq.~\eqref{eq:omGI}. With these two gauge invariants in hand, we may re-write the projections of the \(\ell=0\) EFEs in terms of our gauge invariants. This is done by finding various combinations of Eqs.~\eqref{eq:gaugeinvlzero} to be substituted in the EFE projections found in App.~\ref{sec:einsteina-k}:
\begin{align}
E_\text{K}&=\frac{4}{r^2}\psi_0+\frac{4}{r-2M}\,o_0,\label{eq:EKlzero}\\
\nonumber\\
E_\text{D}&=\frac{4(r-2M)}{r^2}\tder\psi_0,\label{eq:EDlzero}\\
\nonumber\\
E_\text{A}&=-\frac{4(r-2M)^3}{r^4}\rd\psi_0-\frac{4(r+2M)(r-2M)^2}{r^5}\psi_0,\label{eq:EAlzero}\\
\nonumber\\
E_\text{E}&=2\frac{\partial^2}{\partial t^2}\psi_0+\frac{2(r-M)(r-2M)}{r^3}\rd\psi_0+\frac{4M(r-M)}{r^4}\psi_0+2\frac{\partial}{\partial r}o_0+\frac{2(r-M)}{r(r-2M)}\,o_0.\label{eq:EElzero}
\end{align}
Our desire is to have equations for \(o_0\) and \(\psi_0\) in terms of stress-energy sources. We now invert the above expressions to find equations for \(o_0\) and \(\psi_0\). In practical applications we find it simplest to work with as few time derivatives as possible in solving for our gauge invariants. To begin, we use Eq.~\eqref{eq:EKlzero} to solve for \(o_0\),
\begin{align}
\label{eq:omeqn}
o_0&=\frac{r-2M}{4r^2}\left(r^2E_\text{K}-4\psi_0\right),
\end{align}
and substitute this into the three remaining EFE projections. We then use Eq.~\eqref{eq:EAlzero} to solve for the first radial derivative of \(\psi_0\),
\begin{align}
\label{eq:drdpsilzero}
\frac{\partial}{\partial r}\psi_0&=-\frac{(r+2M)}{r(r-2M)}\psi_0-\frac{r^4}{4(r-2M)^3}E_\text{A}.
\end{align}
One might be tempted to substitute Eqs.~\eqref{eq:EDlzero} and \eqref{eq:drdpsilzero} into Eq.~\eqref{eq:EElzero}, in order to solve for \(\psi_0\) directly, however we must recall that Eqs.~\eqref{eq:EKlzero}-\eqref{eq:EElzero} are not independent, but are related through the Bianchi identities. Performing such a substitution will recover a combination of the two even-parity projections of the Bianchi identities and completely eliminate \(\psi_0\) and \(o_0\).

To recover the metric for \(\ell=0\), we first choose a gauge; this must be done with some care, as the \(\ell=0\) reductions of Eqs.~\eqref{eq:AKtoPQRS} do not allow for an entierly unique choice of gauge. We see that, in setting \(\tS=0\) and by only looking at the gauge effects on A, D, E, and K, the E term may be algebraically removed through the gauge vector component R, but the remaining three metric components are only differentially related to the one remaining degree of gauge freedom found in P. One should then remember that simply setting two components of the metric to zero for \(\ell=0\) does not necessarily indicate a unique choice of gauge! For this analysis, we decide to use the choice of gauge (class) introduced by Zerilli~\cite{zerilli}, whereby we set
\[
\mathrm{D}_0=\mathrm{E}_0=0.
\]
Once this choice is made, we recover the two remaining metric projections \(\mathrm{A}_0\) and \(\mathrm{K}_0\) via Eqs.~\eqref{eq:gaugeinvlzero}. This procedure is worked out in detail in Sec.~\ref{sec:l0example} for the example of particle in circular orbit around a Schwarzschild black hole. 

\subsection{Special Case: \(\ell=1\)}
\label{sec:ellone}
When considering the special case of \(\ell=1\), both the even- and odd-parity sectors must be examined. We perform the analysis of the odd-parity sector first.

\textit{Odd-Parity:} In the odd-parity sector, the transverse-traceless projection of the EFE, \(E_\text{G}\), vanishes identically and we are left with two remaining components of the metric perturbation, \(\text{C}_1\) and \(\text{J}_1\), where the subscript of ``1'' reminds us that we are looking at \(\ell=1\) perturbations. We also have one degree of gauge freedom, generated by Q, and we will use this freedom to find a single odd-parity gauge invariant through the method of Sec.~\ref{sec:generalapproach}. But first we note that the gauge invariants listed in Eqs.~\eqref{eq:indepgaugeinvar} do \textit{not} remain invariant for \(\ell=1\)! This is an obvious effect of the non-existence of the G term. We therefore must find combinations of \(\tJ_1\) and \(\tC_1\) alone that form a gauge invariant, if any exist. One hint on how to do this comes from the Cunningham-Price-Moncrief master variable detailed in Eq.~\eqref{eq:Wdef}. This master variable is written in terms of \(\alpha\) and \(\beta\) for \(\ell\ge2\), which are no longer individually gauge invariant at \(\ell=1\). However, the combination of these invariants remains gauge invariant for \(\ell=1\), and written in terms of \(\tC_1\) and \(\tJ_1\) is,
\begin{align}
W_1&=r^3\left(\frac{\partial}{\partial t}\text{J}_1+\frac{\partial}{\partial r}\text{C}_1-\frac{1}{r}\text{C}_1\right).\label{eq:loneW}
\end{align}
\(W_1\) is the simplest form of gauge invariant quantity to be found, i.e., it has the fewest number of derivatives acting on the metric components. 
The two perturbed EFEs may be written in terms of \(W_1\),
\begin{align}
E_\text{J}&=-\frac{1}{r^2(r-2M)}\frac{\partial}{\partial t}W_1,\label{eq:EJlone}\\
\nonumber\\
E_\text{C}&=\frac{r-2M}{r^4}\frac{\partial}{\partial r}W_1,\label{eq:EClone}
\end{align}
and these equations can also be combined into a wave equation, just as in Eq.~\eqref{eq:RWeq},
\begin{equation}
\left(-\frac{r}{r-2M}\frac{\partial^2}{\partial t^2}+\frac{r-2M}{r}\frac{\partial^2}{\partial r^2}+\frac{2M}{r^2}\frac{\partial}{\partial r}\right)W_1=r^3\left(\frac{\partial}{\partial t}E_\tJ+\frac{\partial}{\partial r}E_\tC+\frac 3r E_\tC\right).
\end{equation}

Once we have solved for \(W_1\), we are interested in reconstructing the metric perturbation, and to do this we first must choose a gauge. From Eq.~\eqref{eq:loneW} it appears to be in our best interest to choose a gauge in which either \(\tJ_1\) or \(\tC_1\) vanishes, which will turn Eq.~\eqref{eq:loneW} into a differential equation for either \(\tJ_1\) or \(\tC_1\) alone, with \(W_1\) as a source term. As with the \(\ell=0\) case, any gauge choice that simply eliminates one component of the metric will not necessarily be unique! Still, from Eqs.~\eqref{eq:AKtoPQRS} it is clear that we may choose a class of gauges in which \(\tJ_1=0\). This choice of gauge (class) greatly simplifies  Eq.~\eqref{eq:loneW}, which we then solve for \(\mathrm{C}_1\) to recover the metric projection. We use this procedure for metric reconstruction in Sec.~\ref{sec:l1example} for our simple example.

\textit{Even-Parity:} Searching for gauge invariants in the even-parity sector for \(\ell=1\) turns out to be a more delicate process than one might expect. We notice that all of our gauge invariants listed in Eqs.~\eqref{eq:indepgaugeinvar} actually remain gauge invariant due to the controlling effects of the F term in our decomposition (the change in the F term is algebraically related to S and \textit{only} S, via Eq.~\eqref{eq:Fchange}). As the F term vanishes identically for \(\ell<2\), the gauge invariants found above simply become combinations of metric components which are all changed by the S component of the gauge vector, losing their invariant properties. 

In fact, we found no \textit{simple} gauge invariant combinations of the metric components for \(\ell=1\) in the even-parity sector via the method outlined in Sec.~\ref{sec:generalapproach} \textit{outside of those combinations of metric components found already in the perturbed Einstein equations,} several of which become first order in metric derivatives as we show below! (One might attempt to find gauge invariant fields using the methods for odd-parity \(\ell=1\) above, however the Zerilli master variable is not gauge invariant at \(\ell=1\), and any simple combination of metric components just recovers the projections of the EFEs.) This is not a surprise, however, as we will explain.

In his work on even-parity perturbations of Schwarzschild, Zerilli \cite{zerilli} found that a gauge exists in which the metric perturbation may be set to zero everywhere outside the support of a compact stress-energy. (Zerilli worked on the case of a radially-infalling point particle, but it is easy to generalize the results to any object with compact spatial support.) Thus, one expects any fields constructed from components of the metric perturbation to vanish outside the support of the stress-energy, including any gauge invariant fields one can find. As an example of this fact, one associates the even-parity \(\ell=1\) metric perturbation with the linear momentum of the system \cite{MTW}. We find that we may set this momentum to zero in a particular gauge outside of the local source region (essentially by putting ourselves in the asymptotic rest frame of the center-of-mass of the system). As the linear momentum vanishes in only a particular choice of coordinates, it is \textit{not} a gauge invariant quantity and not attributed to a physical observable of the system. It seems that the only physical observables in this sector are then local quantities directly related to the stress-energy of our source. This ``center-of-momentum'' gauge transformation was also introduced by Detweiler and Poisson \cite{DetweilerPoisson} for the case of a particle in a circular orbit about a Schwarzschild black hole, and we will recover the general gauge transformation below in Section~\ref{sec:simpex} for this case. The lack of even-parity gauge invariant fields for \(\ell=1\) was also noted by Sarbach and Tiglio \cite{SarbachTiglio}, who studied the relationship between the RW and Zerilli master variables and the ADM quantities.

Despite the lack of additional gauge invariants, we are still able to determine the metric components completely; at \(\ell=1\), the transverse-traceless projection of the EFE on the two-sphere, \(E_\text{F}\), vanishes identically, and we may fully describe the metric perturbation using the proper set \(E_\text{A},E_\text{D}\) and \((E_\text{K}+2E_\text{H})\) from the Bianchi identities. Using the gauge choice that Zerilli introduced for \(\ell=1\), we set \(\text{B}_1=\text{E}_1=\text{H}_1=0\) and recover A\(_1\), D\(_1\), and K\(_1\) from the remaining equations,
\begin{align}
E_\text{A}&=-\frac{2(r-2M)^3}{r^4}\rd\text{K}_1-\frac{4(r+M)(r-2M)^2}{r^5}\text{K}_1,\label{eq:EAlone}\\
\nonumber\\
E_\text{D}&=-\frac{2}{r^2}\text{D}_1+\frac{2(r-2M)}{r^2}\tder\text{K}_1,\label{eq:EDlone}\\
\nonumber\\
E_\text{K}+2E_\text{H}&=\frac{2M}{r^3}\text{K}_1+\frac{2}{r-2M}\tder\text{D}_1-\frac{6M}{r(r-2M)^2}\text{A}_1.\label{eq:EKEHlone}
\end{align}
Solving for the metric projections is quite clear. First, Eq.~\eqref{eq:EAlone} is integrated to solve for \(\text{K}_1\). This solution is then used in Eq.~\eqref{eq:EDlone} to solve for \(\text{D}_1\), and these two solutions are subsequently used in Eq.~\eqref{eq:EKEHlone} to solve for \(\text{A}_1\).
The remaining EFE equations \(E_\text{B}\) and \(E_\text{E}\), along with \(E_\text{H}\) and \(E_\text{K}\) individually, may be used to check the consistency of the solution.


\section{A Simple Example}\label{sec:simpex}
The formalism introduced above has proven to be extremely beneficial for the study of the motion of small objects in a Schwarzschild background. To demonstrate the ease with which problems of this nature may be solved using the A--K notation, we work through one of the simplest problems available in gravitational perturbation theory: the motion of a point particle, mass \(\mu\), in circular orbit about a Schwarzschild black hole, mass \(M\).  Our focus will be on finding analytic solutions to this problem for \(\ell=0,1\), and we will outline the general approach to a solution for \(\ell\ge2\), where general analytic solutions do not exist. An analysis relating the low-multipole results detailed in this section to the low-multipole metric perturbation in the Lorenz gauge was performed by Detweiler and Poisson \cite{DetweilerPoisson}.

The stress-energy of the point particle takes the form,
\begin{align}
\label{eq:stressenergy}
T_{ab}&=\mu\int\frac{u_a u_b}{\sqrt{-g^0}}\delta^{(4)}\left[x-z(\tau)\right]\;\mathrm{d}\tau,
\end{align}
where \(u_a\) is the four-velocity of the particle along the worldline \(z(\tau)\), \(g^0\) is the determinant of the background Schwarzschild metric, and \(\delta^{(4)}\) is the four-dimensional Dirac delta function. The specifics of the theory of distributions on curved spacetimes is nicely summarized by Poisson \textit{et al.} \cite{LivingReview}. Specifically for a circular orbit in the equatorial plane, the four-velocity reduces to
\begin{align}
u_{a}&=\left(-\mathcal{E},0,0,\mathcal{J}\right),
\end{align}
with \(\mathcal{E}\) and \(\mathcal{J}\) being constants of the motion corresponding to the specific energy and angular momentum of the particle, respectively,
\begin{align}
\mathcal{E}&=\frac{R-2M}{\sqrt{R(R-3M)}},&\mathcal{J}&=R\,\sqrt{\frac{M}{R-3M}},
\end{align}
and \(R\) is the orbital radius of the particle. One may also define the orbital frequency,
\begin{align}
\nonumber\Omega^2&=\left(\frac{u^\phi}{u^t}\right)^2\\
&=\left(\frac{\mathcal{J}/R^2}{\mathcal{E}/(1-2M/R)}\right)^2\\
\nonumber&=\frac{M}{R^3}.
\end{align}
From Eq.~\eqref{eq:stressenergy} we may calculate the A--K projections of the stress-energy, using the decomposition outlined in App.~\ref{sec:a-kdecomp}, which are then related to the projections of the EFEs via Eq.~\eqref{eq:linearizedproj}. For a stress-energy in the form of Eq.~\eqref{eq:stressenergy}, the D, H, J, and K projections vanish identically, as the four-velocity for a circular orbit has no radial component. We now list the required non-zero projections of the stress-energy, separating by those projections valid for \(\ell\ge0\), \(\ell\ge1\), and \(\ell\ge2\):
\begin{align}
\nonumber\ell\ge0,&\\
&E^{\ell m}_\text{A}=-16\pi\left(1-\frac{2M}{R}\right)\frac{\mu \mathcal{E}}{R^2}\delta\left(r-R\right)Y^{*}_{\ell m}\left(\frac\pi2,\Omega t\right),\label{eq:stressA}\\
&E^{\ell m}_\text{E}=-8\pi\frac{\mu \Omega \mathcal{J}}{R^2}\delta\left(r-R\right)Y^{*}_{\ell m}\left(\frac\pi2,\Omega t\right),\\
\nonumber\ell\ge1,&\\
&E^{\ell m}_\text{B}=\frac{-16\pi}{\ell(\ell+1)}\left(1-\frac{2M}R\right)\frac{\mu \mathcal{J}}{R^3}\delta\left(r-R\right)\left.\left(\frac{\partial}{\partial \phi}Y^{*}_{\ell m}(\theta,\phi)\right)\right|_{\theta=\frac{\pi}{2},\phi=\Omega t},\\
&E^{\ell m}_\text{C}=\frac{-16\pi}{\ell(\ell+1)}\left(1-\frac{2M}R\right)\frac{\mu \mathcal{J}}{R^3}\delta\left(r-R\right)\left.\left(\frac{\partial}{\partial \theta}Y^{*}_{\ell m}(\theta,\phi)\right)\right|_{\theta=\frac{\pi}{2},\phi=\Omega t},\label{eq:stressC}\\
\nonumber\ell\ge2,&\\
&E^{\ell m}_\text{F}=-16\pi\frac{(\ell+2)!}{(\ell-2)!}\frac{\mu\Omega\mathcal{J}}{R^2}\delta\left(r-R\right)\left.\left(2\frac{\partial^2}{\partial\phi^2}Y^{*}_{\ell m}(\theta,\phi)+\ell(\ell+1)Y^{*}_{\ell m}(\theta,\phi)\right)\right|_{\theta=\frac{\pi}{2},\phi=\Omega t}.
\end{align}

\subsection{A Circular Orbit at \(\ell=0\)}
\label{sec:l0example}

At \(\ell=0\) we begin by solving Eq.~\eqref{eq:drdpsilzero} for \(\psi_0\) and use this solution in Eq.~\eqref{eq:omeqn} to recover \(o_0\):
\begin{align}
\psi_0(t,r)&=-\frac{r}{4(r-2M)^2}\int\frac{r'^3E_\text{A}^{00}}{r'-2M}\,\mathrm{d}r',\label{eq:psizero}\\
o_0(t,r)&=\frac{1}{4r(r-2M)}\int\frac{r'^3E_\text{A}^{00}}{r'-2M}\,\mathrm{d}r'.\label{eq:ozero}
\end{align}
Recall that \(E_\text{K}^{\ell m}=0\) for a circular orbit.
Using the gauge choice introduced in Sec.~\ref{sec:ellzero}, from Eqs.~\eqref{eq:gaugeinvlzero} in this gauge we recover the metric projections from \(\psi_0\) and \(o_0\),
\begin{align}
2\psi_0&=\text{K}_0,\label{eq:psitoKzero}\\
2o_0&=\rd\text{A}_0-\frac{2M}{r(r-2M)}\text{A}_0\label{eq:otoAzero}.
\end{align}

Solving for \(\text{K}_0\) and \(\text{A}_0\) from Eqs.~\eqref{eq:psitoKzero} and \eqref{eq:otoAzero} requires the integration of Eqs.~\eqref{eq:psizero} and \eqref{eq:ozero}, which is straightforward to do given the delta function in Eq.~\eqref{eq:stressA}. Choosing the constants of integration such that the perturbation components vanish within the orbit, the resulting A\(_0\) and K\(_0\) take the forms:
\begin{align}
\tK_0&=-\frac{r}{(r-2M)^2}\frac{R^3}{2(R-2M)}\bar{E}^{00}_\text{A}\Theta(r-R)\\
\tA_0&=-\frac{1}{r}\frac{R^3}{2(R-2M)}\bar{E}^{00}_\text{A}\Theta(r-R)
\end{align}
Here, \(\Theta(r-R)\) is the unit step function, and we have introduced the shorthand \(\bar{E}^{00}_\text{A}\) to denote the fully-evaluated coefficient to the delta function in Eq.~\eqref{eq:stressA} evaluated at \(\ell=m=0\),
\[
\bar{E}^{00}_\text{A}=-8\sqrt{\pi}\left(1-\frac{2M}{R}\right)\frac{\mu \mathcal{E}}{R^2}.
\]
With \(\text{A}_0\) and \(\text{K}_0\) now in hand, we may use Eq.~\eqref{eq:hina-k} to recover the metric perturbation. Inside the orbit, the perturbation vanishes. Looking outside the orbit, we recover the metric perturbation,
\[
h^{00}_{ab}=\frac{2\mu \mathcal{E}}{r}v_av_b+\frac{2\mu r \mathcal{E}}{(r-2M)^2}n_an_b.
\]
This perturbation in the Zerilli gauge manifests as a shift to the mass of the system, as seen by adding the perturbation to the background Schwarzschild geometry,
\begin{align*}
g_{tt}&=-\left[1-2\frac{(M+\mu \mathcal{E})}{r}\right],\\
g_{rr}&=\frac{1}{1-2(M+\mu \mathcal{E})/r}+O(\mu^2).
\end{align*}

\subsection{A Circular Orbit at \(\ell=1\)}
\label{sec:l1example}

To perform the analysis at \(\ell=1\), we look separately at the even- and odd-parity sectors.

\textit{Even-Parity:} Unlike for \(\ell=0\), we have found no gauge invariant quantities in the even-parity sector, yet we may still recover the metric perturbation analytically by solving Eqs.~\eqref{eq:EAlone}-\eqref{eq:EKEHlone} directly. We first impose the gauge choice of Zerilli, as outlined in Sec.~\ref{sec:ellone}, by setting \(\text{B}_1=\text{E}_1=\text{H}_1=0\). From there, we solve for the three remaining metric perturbation projections from the \(E_\text{A}^{1m}\) source alone. 
We define the fully-evaluated coefficient of the delta function in Eq.~\eqref{eq:stressA},
\[
\bar{E}^{1m}_\text{A}=-16\pi\left(1-\frac{2M}{R}\right)\frac{\mu\mathcal{E}}{R^2}Y^*_{1m}\left(\frac{\pi}{2},\Omega t\right).
\]
This source term vanishes for \(m=0\), so we must only consider the \(m=\pm1\) contributions to the metric perturbation for even-parity.
Using Eq.~\eqref{eq:EAlone} we may solve for \(K_1\):
\begin{equation}
\tK_1=-\frac{r}{2(r-2M)^3}R^3\bar{E}^{1m}_\text{A}\Theta(r-R).
\end{equation}
We choose the constants of integration such that the perturbation vanishes inside of the orbit. 
The \(\text{D}_1\) term is then trivial to obtain from Eq.~\eqref{eq:EDlone}, 
\begin{equation}
\tD_1=\frac{im\Omega r}{2(r-2M)^2}R^3\bar{E}^{1m}_\text{A}\Theta(r-R).
\end{equation}
The factor of \(-im\Omega\) comes from the time-dependence of \(Y_{1m}^*(\pi/2,\Omega t)\sim e^{-im\Omega t}\) in \(\bar{E}^{1m}_\text{A}\). Finally we recover the \(\tA_1\) term from Eq.~\eqref{eq:EKEHlone},
\begin{equation}
\tA_1=-\frac{R^3}{6r(r-2M)}\bar{E}^{1m}_\text{A}\left(1-\frac{\Omega^2r^3}{M}\right)\Theta(r-R).
\end{equation}
The full metric perturbation is then recovered via Eq.~\eqref{eq:hina-k}.  From our choice of integration constants, the even-parity metric perturbation at \(\ell=1\) vanishes inside of the orbit. As discussed by Zerilli \cite{zerilli}, the even-parity \(\ell=1\) metric perturbation may always be entirely removed via a gauge transformation in a \textit{vacuum} region of the spacetime. For our example, we may set the metric perturbation to zero everywhere \textit{except} at the point \(r=R\), where the perturbation becomes singular.  Detweiler and Poisson \cite{DetweilerPoisson} perform this gauge transformation and produce the resulting singular metric perturbation as a step toward transforming the Zerilli gauge into the Lorenz gauge. In terms of our notation, the gauge vector required will have the components,
\begin{align*}
\tP_1&=\frac{-im\Omega r^2R^3\bar{E}^{1m}_\text{A}}{12M(r-2M)}\,\Theta(r-R),\\
\\
\tR_1&=\frac{r^2R^3\bar{E}^{1m}_\text{A}}{12M(r-2M)^2}\,\Theta(r-R),\\
\\
\tS_1&=\frac{r\bar{E}^{1m}_\text{A}}{12M(r-2M)}\,\Theta(r-R).
\end{align*}

\textit{Odd-Parity:} We solve for the odd-parity \(\ell=1\) metric perturbation through use of the gauge invariant \(W_1\) defined in Sec.~\ref{sec:ellone}. As \(E_\text{J}\) vanishes identically, Eq.~\eqref{eq:EJlone} constrains \(W_1\) to be constant in \(t\) and we thusly consider only solutions with azimuthal number \(m=0\). The fully-evaluated coefficient for the single odd-parity stress-energy term, Eq.~\eqref{eq:stressC}, is
\[
\bar{E}_\text{C}^{10}=-8\pi\left(1-\frac{2M}R\right)\frac{\mu\mathcal{J}}{R^3}\left.\left(\frac{\partial}{\partial \theta}Y^{*}_{10}(\theta,\phi)\right)\right|_{\theta=\frac{\pi}{2},\phi=\Omega t},
\]
and as such, \(W_1\) obeys the equation
\begin{equation}
\rd W_1=\frac{R^4}{R-2M}\bar{E}_\text{C}^{10}\delta\left(r-R\right). \label{eq:gammader}
\end{equation}
Using the gauge choice introduced in Sec.~\ref{sec:ellone}, we see the expression for \(W_1\) simplifies to,
\begin{equation}
W_1 = r^4\rd\left(\frac1r\text{C}_1\right),\label{eq:gammaingauge}
\end{equation}
and Eqs.~\eqref{eq:gammader} and \eqref{eq:gammaingauge} then give us a solution for the metric perturbation directly,
\begin{equation}
\rd\left[ r^4\rd\left(\frac1r\text{C}_1\right)\right]=\frac{R^4}{R-2M}\bar{E}_\text{C}^{10}\delta\left(r-R\right).
\end{equation}
The general solution contains two arbitrary constants of integration, \(f_1\) and \(f_2\),
\[
\text{C}_1=rf_1+\frac1{r^2}f_2+\frac{rR}{3(R-2M)}\bar{E}_\text{C}^{10}\Theta(r-R)-\frac{R^4}{3r^2(R-2M)}\bar{E}_\text{C}^{10}\Theta(r-R).
\]
We may set \(f_2\) to zero, as the background Schwarzschild black hole has no intrinsic angular momentum. The constant \(f_1\) is chosen to eliminate the linear-in-\(r\) piece of the particular solution after we cross the orbit,
\[
f_1=-\frac{R}{3(R-2M)}\bar{E}_\text{C}^{10},
\]
to ensure that the solution remains well-behaved at infinity. The final result is
\begin{equation}
\label{eq:c1sol}
\text{C}_1=-\frac{rR}{3(R-2M)}\bar{E}_\text{C}^{10}+\frac{rR}{3(R-2M)}\bar{E}_\text{C}^{10}\Theta(r-R)-\frac{R^4}{3r^2(R-2M)}\bar{E}_\text{C}^{10}\Theta(r-R).
\end{equation}
The metric perturbation is recovered via Eq.~\eqref{eq:hina-k},
\begin{equation}
h_{t\phi}^{10}=- \text{C}_1\,r\sin\theta\,\frac{\partial}{\partial \theta}Y_{10}(\theta,\phi),
\end{equation}
which, when combined with Eq.~\eqref{eq:c1sol}, yields a metric perturbation which is continuously matched over the orbit of the particle,
\begin{equation}
\label{eq:htpl1}
h_{t\phi}^{10}=-2\mu\mathcal{J}\sin^2\theta\left[\frac{r^2}{R^3}\Theta(R-r)+\frac1r\Theta(r-R)\right].
\end{equation}
Inside the orbit, the contribution to \(h_{t\phi}\) may be seen as pure gauge; in other words, the introduction of the gauge vector \(\tilde\xi_a(t,r)=\tilde Q(t,r)Y_a^{B,10}\) with explicit time dependence \(\tilde Q(t,r)=4\sqrt{\frac{\pi}{3}}\frac{r\mu\mathcal{J}}{R^3}t\) would eliminate \(\tC_1\) inside the orbit without affecting the gauge choice \(\tJ_1=0\) (this may be thought of as a refinement of the original Zerilli gauge, which was not uniquely determined by the requirement that \(\tJ_1\) vanish). In this gauge, however, the metric perturbation would contain a jump discontinuity at the location of the particle, and it would grow quadratically with \(r\) outside of the orbital radius of the particle leading to an ill-behaved metric perturbation at spatial infinity.

Returning to the original Zerilli gauge used to derive Eq.~\eqref{eq:htpl1}, we see that outside of the orbit the specific angular momentum of the particle contributes to the total angular momentum of the system, so that the system now appears to be that of a black hole with a small amount of angular momentum.

\subsection{A Circular Orbit for \(\ell\ge2\)}
\label{sec:l2example}
The analysis performed above for \(\ell=0,1\) benefits from the fact that, with the choice of gauge introduced by Zerilli for even-parity and setting \(\tJ_1=0\) for odd-parity, the metric perturbation may be found analytically. Sadly, this is not the case for the general problem at \(\ell\ge2\). Here we sketch an outline of the solution method typically employed to solve for the \(\l\ge2\) metric perturbation. 

We begin by moving the problem into the frequency domain; the sources for Eqs.~\eqref{eq:zerillieq} and \eqref{eq:RWeq} both exhibit explicit time-dependence as \(\sim e^{-im\Omega t}\), which reduces the number of Fourier modes of our master variables to a finite set for each \(\ell\), with frequencies quantized by the azimuthal index \(m\), \(\omega_m=m\Omega\). The differential equations governing the Fourier modes of the master variables then become second-order ordinary differential equations in \(r\) with distributional sources that vanish almost everywhere (except at the location of the particle, \(r=R\)). 

To integrate these equations, we first split the domain of integration into two parts: a region bounded by the event horizon of the black hole and the orbital radius of the particle, \(r_-\in(2M,R)\), and a region bounded by the location of the particle and spatial infinity, \(r_+\in(R,\infty)\). The method of variation of parameters \cite{BenderOrzag} is employed to integrate the homogeneous equations in each of these domains separately while enforcing appropriate radiative boundary conditions at the event horizon and at spatial infinity. Finally, the two solutions are matched at the particle's orbital radius via the jump conditions generated by the source terms that exist there. While this integration is done numerically for \(m\ne 0\), analytic solutions to the \(m=0\) homogeneous Regge-Wheeler and Zerilli equations satisfying the appropriate boundary conditions exist in terms of Hypergeometric functions (see the Appendix of Field \textit{et al.} \cite{Field} for an outline of the solutions).

With the master variables determined, the gauge invariant quantities may be recovered via Eqs.~\eqref{eq:chifromZ}-\eqref{eq:psifromZ} and \eqref{eq:deltaGIsol}-\eqref{eq:epsGIsol} for the even-parity sector and Eqs.~\eqref{eq:alpha}-\eqref{eq:beta} for the odd-parity sector. Metric reconstruction then follows via Eqs.~\eqref{eq:indepgaugeinvar} after a choice of gauge is made. The beauty of this framework is in the fact that choice of gauge is made only at the very end of the calculation; whether one wishes to investigate e.g., local behavior of the metric perturbation at the location of the particle in the RW gauge or asymptotic behaviors in an asymptotically-flat gauge, the majority of the work may be recycled and only the determination of gauge and metric reconstruction is changed.

For a full picture of the \(\ell\ge2\) solution, we suggest that the reader investigate several sources in the literature: for a numerical time-domain analysis, we recommend the work of Martel \cite{Martel}. For an expanded view on the frequency-domain analysis outlined above, we suggest Cutler \textit{et al.}~\cite{CutlerKennefickPoisson} and Hopper and Evans \cite{HopperEvans}. Additionally, much work has been done to solve the problem of a particle in a circular (or more generally, bound) orbit for \(\ell\ge2\) outside of the Regge-Wheeler and Zerilli formalism, and we strongly recommend that the reader investigate the works of Barack and Lousto~\cite{BarackLousto} and Akcay \textit{et al.}~\cite{AkcayWarburtonBarack} for a specific approach to solving the bound-orbit problem in the Lorenz gauge. Results have been compared between these two approaches within the scope of the gravitational self-force problem \cite{SagoBarackDetweiler}.

\section{Discussion}\label{sec:disc}
Steven Detweiler has written a number of major papers on the self force problem using the approach we have outlined above for considering metric perturbations of the Schwarzschild spacetime.  What we have shown has been compiled from notes principally in the form of Maple code and output, augmented by our own understanding where necessary. Clearly, it already represents a very valuable working tool and we present it here so that it may be of use to the wider community, and specifically so that it may accelerate the analysis of self force problems.    

In principle, the work shown here can be used to compare the results of different computations in greater detail than was previously possible.  For example, if one gauge invariant is plotted against another, some universal curve should result which is unchanged, irrespective of the specific calculation used to obtain it.  Furthermore, these invariants should allow for a much richer and more informative comparison with post-Newtonian and Effective-One-Body (EOB) results.  Such comparisons remain as work for the future.

\section{Acknowledgments}\label{sec:ackn}

The authors are grateful for the helpful and constructive comments of the journal referees.  The authors acknowledge support from NSF grants PHY-1205906, PHY-1314529 and PHY-1607323.  HC acknowledges support through a Summer RA from the IFT at UF.

\vfill
\pagebreak

\begin{appendices}

\section{A-K Decomposition}
\label{sec:a-kdecomp}

The normalizations functions outlined in Sec.~\ref{sec:harmonicbasis} may be found by projecting each vector and tensor harmonic into itself over the 2-sphere:

\begin{multicols}{2}
\begin{align*}
\oint Y_a^{E,\ell m}\left(Y^a_{E,\ell' m'}\right)^*\,\mathrm{d}\Omega&=\ell(\ell+1)\delta_{\ell\ell'}\,\delta_{mm'},\\
\\
\oint Y_a^{B,\ell m}\left(Y^a_{B,\ell' m'}\right)^*\,\mathrm{d}\Omega&=\ell(\ell+1)\delta_{\ell\ell'}\,\delta_{mm'},\\
\\
\oint Y_a^{R,\ell m}\left(Y^a_{R,\ell' m'}\right)^*\,\mathrm{d}\Omega&=\left(1-\frac{2M}{r}\right)\delta_{\ell\ell'}\delta_{mm'},\\
\end{align*}

\begin{align*}
\oint T^{T0,\ell m}_{ab}\left(T_{T0,\ell' m'}^{ab}\right)^*\,\mathrm{d}\Omega&=2\,\delta_{\ell\ell'}\,\delta_{mm'},\\
\\
\oint T^{L0,\ell m}_{ab}\left(T_{L0,\ell' m'}^{ab}\right)^*\,\mathrm{d}\Omega&=\left(1-\frac{2M}{r}\right)^2\delta_{\ell\ell'}\,\delta_{mm'},\\
\\
\oint T^{E1,\ell m}_{ab}\left(T_{E1,\ell' m'}^{ab}\right)^*\,\mathrm{d}\Omega&=\left(1-\frac{2M}{r}\right)\frac{\ell(\ell+1)}{2}\delta_{\ell\ell'}\,\delta_{mm'},\\
\\
\oint T^{B1,\ell m}_{ab}\left(T_{B1,\ell' m'}^{ab}\right)^*\,\mathrm{d}\Omega&=\left(1-\frac{2M}{r}\right)\frac{\ell(\ell+1)}{2}\delta_{\ell\ell'}\,\delta_{mm'},\\
\\
\oint T^{E2,\ell m}_{ab}\left(T_{E2,\ell' m'}^{ab}\right)^*\,\mathrm{d}\Omega&=\frac{(\ell+2)!}{2(\ell-2)!}\delta_{\ell\ell'}\,\delta_{mm'},\\
\\
\oint T^{B2,\ell m}_{ab}\left(T_{B2,\ell' m'}^{ab}\right)^*\,\mathrm{d}\Omega&=\frac{(\ell+2)!}{2(\ell-2)!}\delta_{\ell\ell'}\,\delta_{mm'},\\
\end{align*}
\end{multicols}
The integration is performed over the solid angle \(\mathrm{d}\Omega=\sin\theta\mathrm\,{d}\theta\mathrm{d}\phi\), and the asterisk denotes complex conjugation.

Given a generic tensor perturbation \(h_{ab}\) with an assumed decomposition outlined in Eq.~\eqref{eq:hina-k}, then the A-K components of \(h_{ab}\) may be recovered by projecting onto each associated vector and tensor harmonic. For a particular \(\ell\) and \(m\), and including the required normalizations, we have:

\begin{multicols}{2}
\begin{align*}
\mathrm{A}&=\left(1-\frac{2M}{r}\right)^2\oint v^av^bh_{ab}Y^*_{\ell m}\,\mathrm{d}\Omega,\\
\\
\mathrm{B}&=\frac{-\left(1-\frac{2M}{r}\right)}{\ell(\ell+1)}\oint v^a Y_E^{b*}h_{ab}\,\mathrm{d}\Omega,\\
\\
\mathrm{C}&=\frac{-\left(1-\frac{2M}{r}\right)}{\ell(\ell+1)}\oint v^a Y_B^{b*}h_{ab}\,\mathrm{d}\Omega,\\
\\
\mathrm{D}&=-\oint v^a Y_R^{b*}h_{ab}\,\mathrm{d}\Omega,\\
\\
\mathrm{E}&=\frac{1}{2}\oint T_{T0}^{ab*}h_{ab}\,\mathrm{d}\Omega,\\
\end{align*}
 
\begin{align*}
\mathrm{F}&=\frac{2(l-2)!}{(\ell+2)!}\oint T_{E2}^{ab*}h_{ab}\,\mathrm{d}\Omega,\\
\\
\mathrm{G}&=\frac{2(l-2)!}{(\ell+2)!}\oint T_{B2}^{ab*}h_{ab}\,\mathrm{d}\Omega,\\
\\
\mathrm{H}&=\frac{(l-1)!}{(\ell+1)!}\left(1-\frac{2M}{r}\right)^{-1}\oint T_{E1}^{ab*}h_{ab}\,\mathrm{d}\Omega,\\
\\
\mathrm{J}&=\frac{(l-1)!}{(\ell+1)!}\left(1-\frac{2M}{r}\right)^{-1}\oint T_{B1}^{ab*}h_{ab}\,\mathrm{d}\Omega,\\
\\
\mathrm{K}&=\left(1-\frac{2M}{r}\right)^{-2}\oint T_{L0}^{ab*}h_{ab}\,\mathrm{d}\Omega.\\
\end{align*}
\end{multicols}
This notation is expanded upon in Sec.~\ref{sec:a-kproj} for any rank 2 tensor.

To address the apparent disorder of this decomposition when viewed in only the even- or odd-parity sectors, as mentioned at the end of Sec.~\ref{subsec:evenRWcompare}, we make note of the relationship between the even- and odd-parity vector and tensor harmonics of the same type. As an example, we look at the B and C terms of the metric perturbation and rewrite the pure-spin vector harmonics \(Y^E_a\) and \(Y^B_a\) in terms of spin-weighted spherical harmonics \({}_sY^{\ell m}\)  with spin-weight \(s\) (see Thorne \cite{Thorne}) ignoring normalization factors,
\begin{align*}
\tB v_{(a}Y_{b)}^{E,\ell m}+\tC v_{(a}Y_{b)}^{B,\ell m}&= v_{(a}\left[\tB\left({}_{-1}Y^{\ell m}\mathrm{m}_{b)}-{}_{1}Y^{\ell m}\mathrm{m}^*_{b)}\right)-i\tC\left({}_{-1}Y^{\ell m}\mathrm{m}_{b)}+{}_{1}Y^{\ell m}\mathrm{m}^*_{b)}\right)\right]\\
&=v_{(a}\left[
\left(\tB-i\tC\right){}_{-1}Y^{\ell m}\mathrm{m}_{b)}-
\left(\tB+i\tC\right){}_{1}Y^{\ell m}\mathrm{m}^*_{b)}
\right]\\
&=v_{(a}\left[
\left(\tB-i\tC\right){}_{-1}Y^{\ell m}\mathrm{m}_{b)}+(-1)^m\left\{\left(\tB-i\tC\right){}_{-1}Y^{\ell (-m)}\mathrm{m}_{b)}\right\}^*
\right],
\end{align*}
where \(\mathrm{m}=\frac{1}{\sqrt{2}}(e_\theta+ie_\phi)\) is the complex Newman-Penrose basis vector on the 2-sphere. We see that B and C may be thought of as the real and imaginary parts of a complex function that determines the  behavior of the metric perturbation components \(h_{t\theta}\) and \(h_{t\phi}\) at spin-weight \(s=\pm1\). The same relationship holds for H and J controlling the \(h_{r\theta}\) and \(h_{r\phi}\) components of the metric perturbation (also related via spin-weight \(s=\pm1\) harmonics), and F and G for the transverse-traceless 2-sphere components, which are related to the spin-weight \(s=\pm2\) harmonics.

\newpage
\section{A-K Projections of Linearized Einstein Tensor}
\label{sec:einsteina-k}
It is computationally beneficial to decompose the linearized Einstein tensor, defined for a tensor perturbation \(h_{ab}\) as \(E_{ab}\equiv-16\pi T_{ab}\), into its A-K components. The following expressions are all evaluated generally at \(\ell\geq2\).

\subsection{Even-Parity}

\begin{align*}
E_\text{A}&
=\frac{2(r-2M)^2}{r^2}\edt{r}{E}
+\frac{2(r-2M)(3r-5M)}{r^3}\ed{r}{E}
-\frac{(r-2M)(\ell+2)(\ell-1)}{r^3}\text{E}\\
\\
&\,\,\,\,\,\,-\frac{\ell(\ell+2)(\ell+1)(\ell-1)(r-2M)}{2r^3}\text{F}
+\frac{2\ell(\ell+1)(r-2M)^2}{r^3}\ed{r}{H}
+\frac{2\ell(\ell+1)(r-2M)(2r-3M)}{r^4}\text{H}\\
\\
&\,\,\,\,\,\,-\frac{2(r-2M)^3}{r^4}\ed{r}{K}
-\frac{(2r+4M+r\ell+r\ell^2)(r-2M)^2}{r^5}\text{K}
\end{align*}

\begin{align*}
E_\text{B}&
=\frac{r-2M}{r}\edt{r}{B}
+\frac{2(r-2M)}{r^2}\ed{r}{B}
+\frac{4M}{r^3}\text{B}
-\frac{r-2M}{r^2}\ed{r}{D}
-\frac{2M}{r^3}\text{D}
-\frac{1}{r}\ed{t}{E}\\
\\
&\,\,\,\,\,\,-\frac{(\ell+2)(\ell-1)}{2r}\ed{t}{F}
+\frac{r-2M}{r}\edm{t}{r}{H}
+\frac{3(r-2M)}{r^2}\ed{t}{H}
-\frac{r-2M}{r^2}\ed{t}{K}
\end{align*}

\begin{align*}
E_\text{D}&
=\frac{\ell(\ell+1)}{r}\ed{r}{B}
+\frac{\ell(\ell+1)(r-4M)}{r^2(r-2M)}\text{B}
-\frac{\ell(\ell+1)}{r^2}\text{D}
-2\edm{t}{r}{E}\\
\\
&\,\,\,\,\,\,-\frac{2(r-3M)}{r(r-2M)}\ed{t}{E}
-\frac{\ell(\ell+1)}{r}\ed{t}{H}
+\frac{2(r-2M)}{r^2}\ed{t}{K}
\end{align*}

\begin{align*}
E_\text{E}&
=\edt{r}{A}
+\frac{r-3M}{r(r-2M)}\ed{r}{A}
-\frac{4M^2-2(\ell^2+\ell+2)Mr+\ell(\ell+1)r^2}{2r^2(r-2M)^2}\text{A}
-\frac{\ell(\ell+1)}{r-2M}\ed{t}{B}\\
\\
&\,\,\,\,\,\,+2\edm{t}{r}{D}
+\frac{2(r-M)}{r(r-2M)}\ed{t}{D}
-\frac{r-2M}{r}\edt{r}{E}
+\frac{r}{r-2M}\edt{t}{E}
-\frac{2(r-M)}{r^2}\ed{r}{E}\\
\\
&\,\,\,\,\,\,-\frac{\ell(\ell+1)(r-2M)}{r^2}\ed{r}{H}
-\frac{\ell(\ell+1)}{r^2}\text{H}
+\edt{t}{K}
+\frac{(r-2M)(r-M)}{r^3}\ed{r}{K}\\
\\
&\,\,\,\,\,\,-\frac{4M^2+2(\ell^2+\ell-2)Mr-\ell(\ell+1)r^2}{2r^4}\text{K}
\end{align*}

\begin{align*}
E_\text{F}&
=-\frac{1}{r(r-2M)}\text{A}
-\frac{2}{r-2M}\ed{t}{B}
+\frac{r-2M}{r}\edt{r}{F}
-\frac{r}{r-2M}\edt{t}{F}
+\frac{2(r-M)}{r^2}\ed{r}{F}\\
\\
&\,\,\,\,\,\,-\frac{2(r-2M)}{r^2}\ed{r}{H}
-\frac{2}{r^2}\text{H}
+\frac{r-2M}{r^3}\text{K}
\end{align*}

\begin{align*}
E_\text{H}&
=-\frac{1}{r-2M}\ed{r}{A}
+\frac{r-M}{r(r-2M)^2}\text{A}
-\frac{r}{r-2M}\edm{t}{r}{B}
+\frac{1}{r-2M}\ed{t}{B}
-\frac{1}{r-2M}\ed{t}{D}\\
\\
&\,\,\,\,\,\,+\frac1r\ed{r}{E}
+\frac{(\ell+2)(\ell-1)}{2r}\ed{r}{F}
-\frac{r}{r-2M}\edt{t}{H}
+\frac{2}{r^2}\text{H}
-\frac{r-M}{r^3}\text{K}
\end{align*}

\begin{align*}
E_\text{K}&
=\frac{2}{r-2M}\ed{r}{A}
-\frac{4M+\ell(\ell+1)r}{r(r-2M)^2}\text{A}
-\frac{2\ell(\ell+1)r}{(r-2M)^2}\ed{t}{B}
+\frac{4}{r-2M}\ed{t}{D}
+\frac{2r^2}{(r-2M)^2}\edt{t}{E}\\
\\
&\,\,\,\,\,\,-\frac{2(r-M)}{r(r-2M)}\ed{r}{E}
+\frac{(\ell+2)(\ell-1)}{r(r-2M)}\text{E}
+\frac{\ell(\ell+2)(\ell+1)(\ell-1)}{2r(r-2M)}\text{F}
-\frac{2\ell(\ell+1)(r-M)}{r^2(r-2M)}\text{H}
+\frac{2}{r^2}\text{K}
\end{align*}

\subsection{Odd-Parity}

\begin{align*}
E_\text{C}&
=\frac{r-2M}{r}\edt{r}{C}
+\frac{2(r-2M)}{r^2}\ed{r}{C}
-\frac{\ell(\ell+1)r-4M}{r^3}\text{C}
-\frac{(\ell+2)(\ell-1)}{2r}\ed{t}{G}\\
\\
&\,\,\,\,\,\,+\frac{r-2M}{r}\edm{t}{r}{J}
+\frac{3(r-2M)}{r^2}\ed{t}{J}
\end{align*}

\begin{align*}
E_\text{G}&
=-\frac{2}{r-2M}\ed{t}{C}
+\frac{r-2M}{r}\edt{r}{G}
-\frac{r}{r-2M}\edt{t}{G}
+\frac{2(r-M)}{r^2}\ed{r}{G}\\
\\
&\,\,\,\,\,\,-\frac{2(r-2M)}{r^2}\ed{r}{J}
-\frac{2}{r^2}\text{J}
\end{align*}

\begin{align*}
E_\text{J}&
=-\frac{r}{r-2M}\edm{t}{r}{C}
+\frac{1}{r-2M}\ed{t}{C}
+\frac{(\ell+2)(\ell-1)}{2r}\ed{r}{G}
-\frac{r}{r-2M}\edt{t}{J}\\
\\
&\,\,\,\,\,\,-\frac{(\ell+2)(\ell-1)}{r^2}\text{J}
\end{align*}
\vfill

\pagebreak
\section{A-K Projections of Linearized Einstein Tensor in Terms of Gauge Invariants}
\label{sec:einsteina-kgauge}

\subsection{Even-Parity}
\label{sec:a-kgaugeeven}
\begin{align*}
E_\text{A}&
=\frac{2(r-2M)^2\ell(\ell+1)}{r^3}\rd\chi
+\frac{2(r-2M)(2r-3M)\ell(\ell+1)}{r^4}\chi
-\frac{2(r-2M)^2[4M+2r+r\ell(\ell+1)]}{r^5}\psi\\
&\qquad
-\frac{4(r-2M)^3}{r^4}\rd\psi
\end{align*}

\begin{align*}
E_\text{B}&
=-\frac{2(r-2M)}{r^2}\tder\psi
+\frac{r-2M}{r}\dtdr\chi
+\frac{3(r-2M)}{r^2}\tder\chi
-\frac{r-2M}{r^2}\rd\delta
-\frac{2M}{r^3}\delta
\end{align*}

\begin{align*}
E_\text{D}&
=-\frac{\ell(\ell+1)}{r^2}\delta
+\frac{4(r-2M)}{r^2}\tder\psi
-\frac{\ell(\ell+1)}{r}\tder\chi
\end{align*}

\begin{align*}
E_\text{E}&
=2\tdd\psi
+\frac{2(r-2M)(r-M)}{r^3}\rd\psi
+\frac{4M(r-M)+r(r-2M)\ell(\ell+1)}{r^4}\psi\\
&\qquad-\frac{(r-2M)\ell(\ell+1)}{r^2}\rd\chi
-\frac{\ell(\ell+1)}{r^2}\chi
+2\dtdr\delta
+\frac{2(r-M)}{r(r-2M)}\tder\delta
-2\rdd\epsilon
-\frac{2(r-3M)}{r(r-2M)}\rd\epsilon\\
&\qquad+\frac{r(r-2M)\ell(\ell+1)-4M(r-M)}{r^2(r-2M)^2}\epsilon
\end{align*}

\begin{align*}
E_\text{F}&
=\frac{2}{r(r-2M)}\epsilon
+\frac{2(r-2M)}{r^3}\psi
-\frac{2}{r^2}\chi
-\frac{2(r-2M)}{r^2}\rd\chi
\end{align*}

\begin{align*}
E_\text{H}&
=-\frac{2(r-M)}{r^3}\psi
-\frac{r}{r-2M}\tdd\chi
+\frac{2}{r^2}\chi
-\frac{1}{r-2M}\tder\delta
+\frac{2}{r-2M}\rd\epsilon
-\frac{2(r-M)}{r(r-2M)^2}\epsilon
\end{align*}

\begin{align*}
E_\text{K}&
=\frac{4}{r^2}\psi
-\frac{2(r-M)\ell(\ell+1)}{r^2(r-2M)}\chi
+\frac{4}{r-2M}\tder\delta
-\frac{4}{r-2M}\rd\epsilon
+\frac{2[4M+r\ell(\ell+1)]}{r(r-2M)^2}\epsilon
\end{align*}

\subsection{Odd-Parity}
\label{sec:a-kgaugeodd}

\begin{align*}
E_\text{C}&
=\frac{r-2M}{r}\dtdr\alpha
+\frac{3(r-2M)}{r^2}\tder\alpha
-\frac{r-2M}{r}\rdd\beta
-\frac{2(r-2M)}{r^2}\rd\beta
+\frac{r\ell(\ell+1)-4M}{r^3}\beta
\end{align*}

\begin{align*}
E_\text{G}&
=\frac{2}{r-2M}\tder\beta
-\frac{2}{r^2}\alpha
-\frac{2(r-2M)}{r^2}\rd\alpha
\end{align*}

\begin{align*}
E_\text{J}&
=-\frac{r}{r-2M}\tdd\alpha
-\frac{(\ell+2)(\ell-1)}{r^2}\alpha
+\frac{r}{r-2M}\dtdr\beta
-\frac{1}{r-2M}\tder\beta
\end{align*}

\subsection{Zerilli Equation Source}
The source to Eq.~\eqref{eq:zerillieq} is,
\begin{align}
\label{eq:zerillieqsource}
S_{Z}&=\frac{r^2}{2(r-2M)(6M+\lambda r)}\left[\frac{r[\lambda(\lambda-2)r^2+2M(7\lambda-18)r+96M^2]}{2(r-2M)(6M+\lambda r)}E_\tA-r^2\frac{\partial}{\partial r}E_\tA-r^2\frac{\partial}{\partial t}E_\tD\right.\\
\nonumber&\qquad\qquad\qquad\qquad\qquad\qquad\left.+(\lambda+2)\left(\frac{6M+\lambda r}{2}E_\tF -(r-2M)E_\tH-\frac{r-2M}{2}E_\tK \right)\right].
\end{align}

\end{appendices}

\end{document}